# Discovering the Sky at the Longest Wavelengths with Small Satellite Constellations

Xuelei Chen[1], Jack Burns[2], Leon Koopmans[3], Hanna Rothkaehi[4], Joseph Silk[5,6], Ji Wu[7], Albert-Jan Boonstra[8], Baptiste Cecconi[9], Cynthia H. Chiang[10,11], Linjie Chen[1], Li Deng[7], Maurizio Falanga[12], Heino Falcke[13], Quanlin Fan[7], Guangyou Fang[14], Anastasia Fialkov[15], Leonid Gurvits[16,17], Yicai Ji[14], Justin C. Kasper[18], Kejia Li[19], Yi Mao[20], Benjamin Mckinley[21], Raul Monsalve[10.22.23], Jeffery B. Peterson[24], Jinsong Ping[1], Ravi Subrahmanyan[25], Harish Vedantham[3], Marc Klein Wolt[13], Fengquan Wu[1], Yidong Xu[1], Jingye Yan[7], Bin Yue[1]

1. National Astronomical Observatories, Chinese Academy of Science, Beijing, 100101, China
2. Department of Astrophysical and Planetary Sciences, University of Colorado at Boulder, Boulder, CO 80309, USA
3. Kapteyn Astronomical Institute, University of Groningen, Groningen, The Netherlands
4. Space Research Centre, Polish Academy of Sciences, Warsaw, Poland
5. Department of Physics and Astronomy, Johns Hopkins University, Baltimore, MD 21218, USA
6. CNRS/UPMC Institut d'Astrophysique de Paris, Paris, France
7. National Space Science Centre, Chinese Academy of Science, Beijing, 100190, China
8. Netherlands Institute for Radio Astronomy (ASTRON), Dwingeloo, Netherlands
9. LESIA, Observatoire de Paris, Meudon, France
10. Department of Physics, McGill University, Montreal, Quebec H3A 2T8, Canada
11. School of Mathematics, Statistics & Computer Science, University of KwaZulu-Natal, Durban, South Africa
12. International Space Science Institute (ISSI), Bern, Switzerland
13. Astronomy Department, Radboud University, Nijmegen, The Netherlands
14. Institute of Electronics, Chinese Academy of Science, Beijing, 100190, China
15. Department of Physics and Astronomy, University of Sussex, Brighton BN1 9QH, UK
16. Joint Institute for VLBI ERIC (JIVE), Dwingeloo, Netherlands
17. Faculty of Aerospace Engineering, Delft University of Technology, Delft, Netherlands
18. Climate and Space Sciences and Engineering, University of Michigan, Ann Arbor, MI 48109, USA
19. Kavli Institute for Astronomy and Astrophysics, Peking University, Beijing 100871, China
20. Department of Astronomy, Tsinghua University, Beijing 100084, China
21. Curtin Institute of Radio Astronomy (CIRA), Curtin University, Perth, Austalia
22. School of Earth and Space Exploration, Arizona State University, Tempe, Arizona 85287, USA
23. Universidad Católica de la Santísima Concepción, Alonso de Ribera, Concepción, Chile.
24. Department of Physics, Carnegie Mellon University, Pittsburgh, PA 15289, USA
25. Raman Research Institute, India

# Abstract

Due to ionosphere absorption and the interference by natural and artificial radio emissions, ground observation of the sky at the decameter or longer is very difficult. This unexplored part of electromagnetic spectrum has the potential of great discoveries, notably in the study of cosmic dark ages and dawn, but also in heliophysics and space weather, planets, cosmic ray and neutrinos, pulsar and interstellar medium, extragalactic radio sources, and even SETI. At a forum organized by the International Space Science Institute-Beijing (ISSI-BJ), we discussed the prospect of opening up this window for astronomical observations by using a constellation of small or micro-satellites. We discussed the past experiments and the current ones such as the low frequency payload on Chang'e-4 mission lander, relay satellite and the Longjiang satellite, and also the future DSL mission, which is a linear array on lunar orbit which can make synthesized map of the whole sky as well as measure the global spectrum. We also discuss the synergy with other experiments, including ground global experiments such as EDGES, SARAS, SCI-HI and High-z, PRIZM/Albatros, ground imaging facillities such as LOFAR and MWA, and space experiments such as SUNRISE, DARE/DAPPER and PRATUSH. We also discussed some technical aspects of the DSL concept.

# 1. Introduction

Over the last century, astronomical observations have expanded from the optical to the radio, infrared, ultraviolet, X-ray and gamma-ray bands of the electromagnetic spectrum, and further supplemented by non-electromagnetic observations such as the cosmic ray, neutrino, and gravitational waves. These new observational domains brought many unexpected discoveries, which greatly changed our view of the Universe, and gave deep insight on the fundamental laws of Nature. However, at the longest wavelengths of the electromagnetic spectrum, our view is still incomplete, as the observations at frequencies below ∼30 MHz are strongly hampered by the ionosphere and human-made radio frequency interferences (RFIs). Even at ∼100 MHz high precision observations can still be affected by these factors. Observations at this unexplored low frequency part of the spectrum might provide unique probes for the dark ages after the Big Bang and Cosmic Dawn when first stars and galaxies formed. It may also shed light on many astrophysical phenomena, from active processes in Sun and planets, through exoplanets, interstellar medium and galactic , to radio galaxies, quasars, clusters and intergalactic medium. They might also reveal previously unknown objects or phenomena. This Forum is dedicated to the exploration of this new observational window of the electromagnetic spectrum.

The Forum reviewed our current understanding on various science subjects related to the low frequency radio window, previous and ongoing observations, recent progress, and key science problems to be solved. It then discussed concepts and technologies related to space-based low frequency radio observation and data processing. In particular, it focused on a possible future lunar orbit array mission of Discovering Sky at the Longest wavelengths (DSL). The proposed array is made up of satellites flying in linear formation, making both interferometric and single antenna observations on the orbit behind the Moon, shielded from the Earth-originated radio frequency interferences. This mission concept is under intensive study by the Chinese Academy of Sciences (CAS) in collaboration with national and international partners.

## 2. The exploration of sky at low frequency

### 2.1 Early ground and space experiments

Radio astronomy was born at what we would call now long wavelengths: the pioneering observations conducted by Karl Jansky in 1932-33 were performed at the wavelength of about 15 m. Further development of radio astronomy was carried out with a strong emphasis on shorter wavelengths. This was dictated by both the astrophysical research agenda and difficulties of astronomy observations at decametric and longer wavelengths due to ionosphere opacity. There were only few noticeable exceptions of radio astronomy facilities that operated at frequencies below ~60 MHz (wavelengths longer than ~5 m). One of them was the Ukrainian T-shaped Radio Telescope (UTR) able to observe at wavelengths up to ~40 m (Braude et al. 1978). At about the same time decametric radio telescopes were put in operation in the USA at Clark Lake (Erickson et al. 1982) and in France at Nançay (Nançay Decameter Array, NDA, Boischot et al. 1980). Some observations below 30MHz were made from ground facilities at Tasmania in the south and Canada in the north (Reber 1994, Bridle & Purton 1968, Caswell 1976, Cane & Whitham 1977, Cane 1978. Roger et al. 1999). Around the turn of the century, the interest to cosmology and astrophysics in the long wavelength spectrum domain stimulated deployment of several large meter-wavelength facilities such as LOFAR[1] (Low Frequency Array) centered in the Netherlands and spread throughout Europe, the LWA[2] (Long Wavelength Array) in the USA, and the MWA[3] (Murchison Widefield Array) in Australia, They all are considered to be pathfinders for the next generation large radio astronomy facility, the SKA (Square Kilometre Array[4]). However, these facilities will not address the strengthening science case for radio astronomy studies at ultra-long-wavelengths, longer than ~10 m. The only solution for addressing the science case of ultra-long wavelength radio astronomy is in placing a telescope beyond the ionosphere, in Space (Jester & Falcke 2009, Boonstra et al. 2016).

---

[1] http://www.lofar.org

[2] http://lwa.unm.edu

[3] http://mwatelescope.org

[4] http://astronomers.skatelescope.org

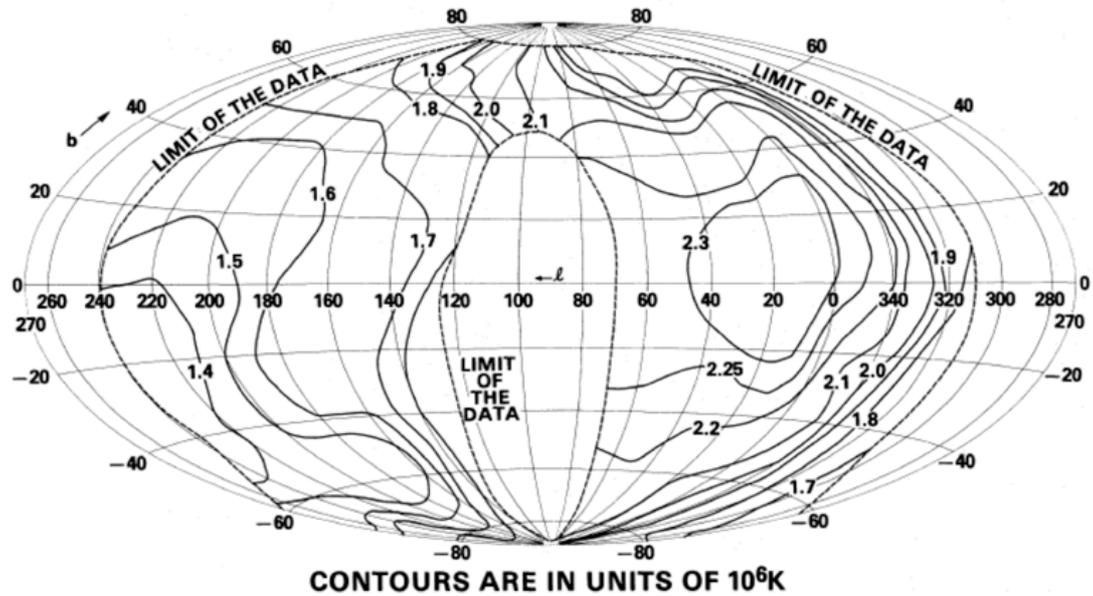

Figure 2-1. The RAE-2 sky map at 4.70 MHz (Novico & Brown 1978).

In the 1970s, the IMP-6 (Brown 1973), the Radio Astronomy Explore (RAE)-1 (Alexander & Novaco 1974) and RAE-2 (Alexander et al. 1975) satellites made low frequency radio observations from space. The data collected by these satellites showed that the Earth have strong natural radio emissions at the kilometric wavelengths, and that man-made radio frequency interferences are visible from space. The Moon can shield the spacecraft from these emissions of the Earth, so the far side of the Moon provides an ideal environment for low frequency radio observation. However, these single antenna observations had poor angular resolution despite a remarkably long antenna deployed in space (Novaco & Brown 1978).

Space-borne radio astronomy is a logical and inevitable step in the overall development of astronomical science. It is driven by three factors: (i) the afore-mentioned opacity of the ionosphere at long and atmosphere at short (millimeter and sub-millimetre) wavelengths, (ii) the ever rising level of human-produced radio frequency interference (RFI), and (iii) the astrophysics-driven demand for higher angular resolution, which in turn requires aperture sizes (i.e., interferometric baselines) larger than the Earth diameter. The latter necessitates extension of the global very long interferometer baselines (VLBI) to orbital dimensions, creating Space VLBI (SVLBI) systems. The history of SVLBI began almost simultaneously with the invention of the VLBI technique as such in the middle of the 1960s. To date, interferometric baselines longer than the Earth diameter produced astrophysical results in the first SVLBI demonstration experiment with the geostationary communication satellite, TDRSS ([Levy et al. 1986](#)), and two first-generation dedicated SVLBI missions, the Japan-led VSOP/HALCA (1997–2003, [Hirabayashi et al. 1998](#)) and Russia-led RadioAstron (2011-2019, [Kardashev et al. 2013)](#). While all three implemented SVLBI systems operated at wavelengths shorter than those under consideration in this Report, lessons learned from their design, construction, tests and operations ([Gurvits 2018, 2019](#)) might be relevant for a prospective ULW spaceborne radio interferometer.

## 2.2 The exploration of the Solar System with space radio astronomy instruments

The main low frequency radio sources in the Solar System are related to Solar activity on one hand, and planetary aurora on the other hand. Many space probe including radio astronomy receivers explored our Solar System. The outer planets (Jupiter, Saturn, Uranus and Neptune) low frequency radio emissions have been discovered by the Voyager PRA (Planetary Radio Astronomy) experiment. The Jupiter and Saturn radio emissions have been extensively studied by the Gallileo/PWS, Cassini/RPWS and Juno/Waves instruments. Solar observations at low frequency have been also studied on the long run with the ISEE3, WIND or STEREO missions. Terrestrial kilometric radiation was observed by many space missions (RAE, Swedish Viking, FAST, Cluster, Interball, Geotail…).

Figure 2-2 shows the normalized power spectral densities for low frequency Solar and Planetary radio emissions, as well as the observation frequency range of the major radio astronomy space probes. The space radio astronomy instruments are limited in sensitivity (single antenna, bright sky background, limited power and downlink rate, limited antenna gain…), but they were developed included so-called "Direction-Finding" or "Goniopolarimetric" capabilities, which allows to retrieve the flux, direction of arrival and polarization of any dominant radio source in the sky (see [Cecconi 2011](#)).

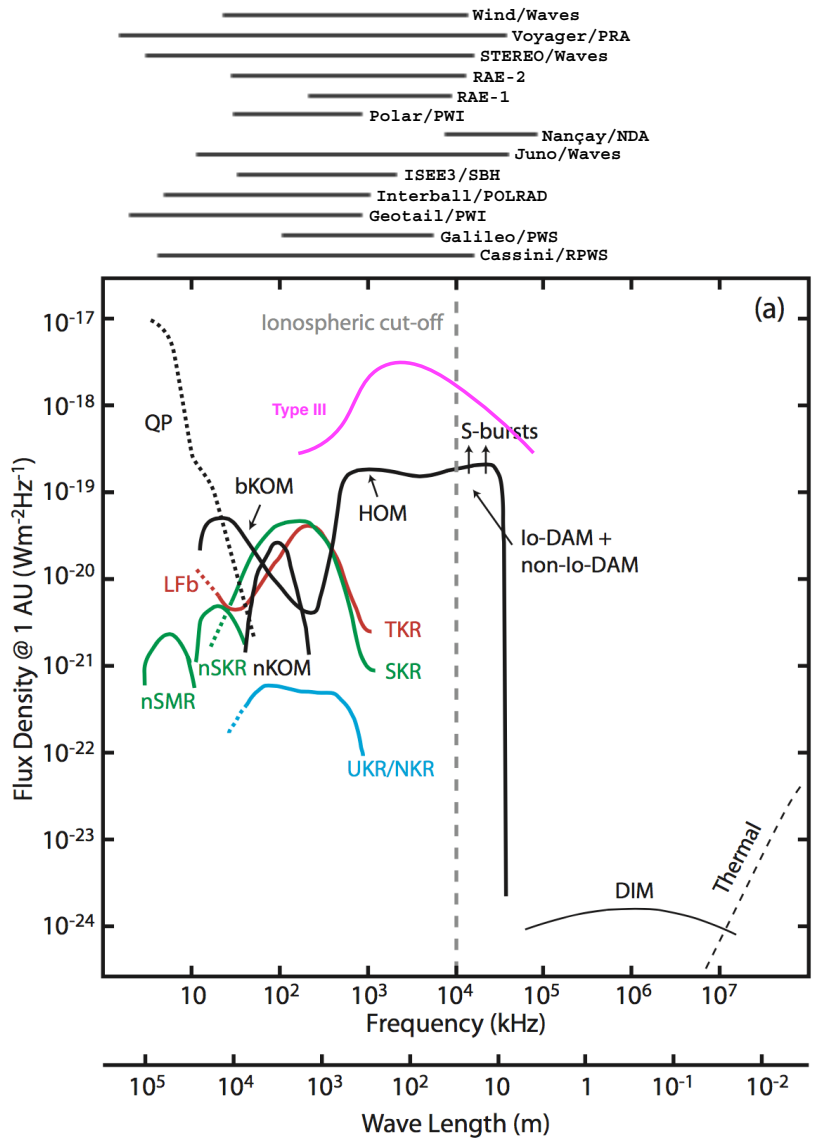

Figure 2-2: Planetary radio emissions (Terrestrial in red, Jupiter in black, Saturn in green, Uranus and Neptune in blue) [extracted from Cecconi, et al 2018] and space probes observation spectral range.

The analysis of data acquired with space-borne radio astronomy instruments with goniopolarimetric capabilities have provided the Solar system science communities with many observational evidences of relativistic accelerated particles populations interacting with a colder ambient plasma, through plasma instabilities (e.g., in the Solar Wind or in the auroral regions of planetary magnetospheres). Low frequency radio astronomy is thus a powerful remote sensing tool for probing energetic distant plasma.

## 2.3 The various mission concepts

Single co-located low-frequency antenna systems are well suited for space science and for studying solar system plasma physics phenomena. Strong celestial signals can even be spatially located, at least for one or a few strong dominating sources as is shown for example in Chang'e 4 (CE4) lunar lander (see Sec.2.4) or Cassini. As at frequencies below about 15 MHz the antenna patterns are spatially symmetrical, a rough indication is needed of the source's origin. For planetary sciences this often is the case. The enabling factor for being able to observe in full polarization and to localize sources is the number of degrees of freedom of the system. For this reason, ideally a spacecraft would have three orthogonal dipoles. If space is limited and only one outer wall of a spacecraft can be used, such as for CE4-RS, then monopoles can be used as well. Spatial nulling of radio interference, either external or self-generated, can in principle also be applied. However, this consumes at least one degree of freedom, so systems should be made flexible in terms of online and offline processing. In this way one could exchange for example optimal polarization performance for interference suppression.

Galactic and extra-galactic radio astronomy science usually requires high spatial resolution to isolate a particular source for research on transient behavior, morphology, polarization, or for spectral analysis. A space-based constellation of small satellites operating in aperture synthesis mode or operating as a phased-array can provide this. Creating sky images are done either by forming several (parallel) broad-band beams, or by creating a full narrow-band cross correlation matrix, and sending these to Earth for further processing and analysis.

The constellation can have fixed relative satellite positions, or the satellites can be slowly drifting, provided their relative positions are known. As with ground-based radio interferometry, the celestial radio signals observed by the satellites are mutually correlated, filling the so-called aperture plane, or aperture sphere for three-dimensional array. As a planar array cannot separate signals coming from the front of the array versus coming from the back, a planar array needs to change it orientation over time. In the DSL concept (Boonstra et al. 2016, Huang et al. 2018) this is achieved by making use of lunar orbit precession.

Different deployment locations of satellite aperture synthesis 'clouds' in space have their pro's and con's. One could deploy them far from Earth's transmitters such as in the Sun-Earth L2 Lagrangian point (e.g. FIRST, Bergman et al., 2009; SURO, Blott et al. 2013). At that location the satellites would only slowly drift, which would allow longer satellite cross-correlation integration times and lower downlink data rates. However, the aperture filling would be very sparse resulting in a somewhat limited instantaneous source separation capability.

In relatively low altitude Lunar orbit a dynamic two or three-dimensional array of satellites such as OLFAR (Engelen et al. 2013), and DSL would have excellent aperture coverage. But due to the fast changing satellite relative positions, the correlation integration times need to be split in very short intervals, resulting in relatively large downlink data rates. Low altitude Lunar orbits are unstable, but putting satellites in the same orbit, as a linear array as in the DSL concept, this issue is circumvented.

A solution that optimizes aperture filling and downlink rates is a constellation in an Earth leading or trailing orbit as proposed in DARIS (Saks et al. 2010). Another concept is the deployment of a rotating tethered string of antennas as proposed in Kruithof et al. (2017). Such a constellation could be combined with one or more free-flyers to provide the third aperture dimension.

## 3. The CE-4 mission

The Chang'e project is an ongoing series of Chinese robotic missions to the Moon. The Chang'e 4 mission put an lander on far side of the Moon for the first time, the landing site is at the Aitken Basin near the lunar south pole. To provide communication link with Earth, a communication relay satellite Queqiao (Magpie Bridge) is launched before the lander to a halo orbit around the Earth-Moon Lagrangian point L2. Taking the opportunity to go to the far side of the Moon, three ultralong wavelength radio projects are carried out: the Longjiang satellites on lunar orbit; the Netherland-China Low-Frequency Explorer on board the Queqiao satellite, and the very low radio frequency spectrometer on the lander.

## 3.1 Longjiang satellites

The Longjiang 1 and 2 satellites were launched into space on May 21, 2018 together with the CE-4 lunar probe's relay satellite. These two satellites were planned to reach the Moon by their own thruster, then form a two element lunar interferometer on orbit to do the interferometric measurement on 1MHz to 30MHz. These two satellites were made by the Harbin Institution of Technology. Longjiang 2 successfully reached its destination near the Moon on May 25, 2018, and entered a lunar orbit with the perilune at 350km and the apolune at 13700km. However, Longjiang 1 suffered an anomaly and failed to enter lunar orbit. The objective for Longjiang 2 was to do spectrum measurement at 1MHz~30MHz, corresponding to the wavelength of 300m to 10m. As illustrated in Figure 3-1, the low frequency interferometer (LFI), was installed on the Longjiang 1 and 2. LFI included deployable antenna, digital receiver, communication ranging and timing synchronization unit.

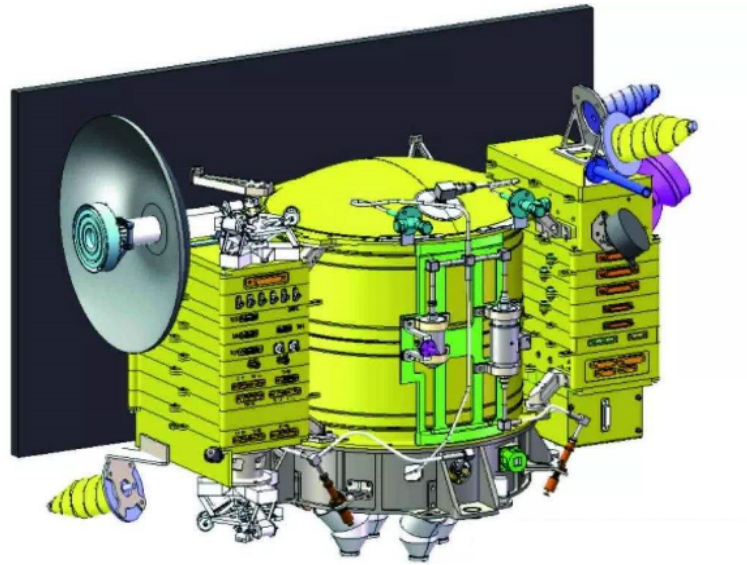

Figure 3-1. Longjiang satellite and LWF (provided by NSSC)

The Upper antenna is made by the Polish Academy of Sciences, and the lower antenna is made by the National Space Science Center, Chinese Academy of Sciences. The low frequency interferometer (LFI) is installed on the Longjiang 1 and 2. The specifications are listed in Table 3-1.

Some key technologies of low frequency interferometer have been validated in this mission, including deployable Antenna, low frequency data receiver, internal calibration and so on. The observational plan were scheduled with the aim of studying radiation characteristic of celestial sources. The Earth occultation experiments and other planets occultation experiments have been scheduled. However, due to limitation of the battery power, the payload worked only 10 to 20 minutes per orbit.

Table 3-1. Specification of Longjiang 2 and LFI

|  | Items | Value |
|---|---|---|
| Longjiang 2 | Weight | 50kg |
|  | Lifetime | 1year |
|  | Orbit | apolune 13700km<br>perilune 350km<br>inclination 36.248deg<br>duration 21hour |
| Low Frequency Interferometer | Weight | 2kg |
|  | Power | 15w |
|  | Antenna | Triple dipole<br>Length 1m |
|  | Frequency range | 1MHz~30MHz |
|  | Channel number | 3 |
|  | Spectrum resolution | <0.1MHz |
|  | Time synchronization precision | 3.3ns |

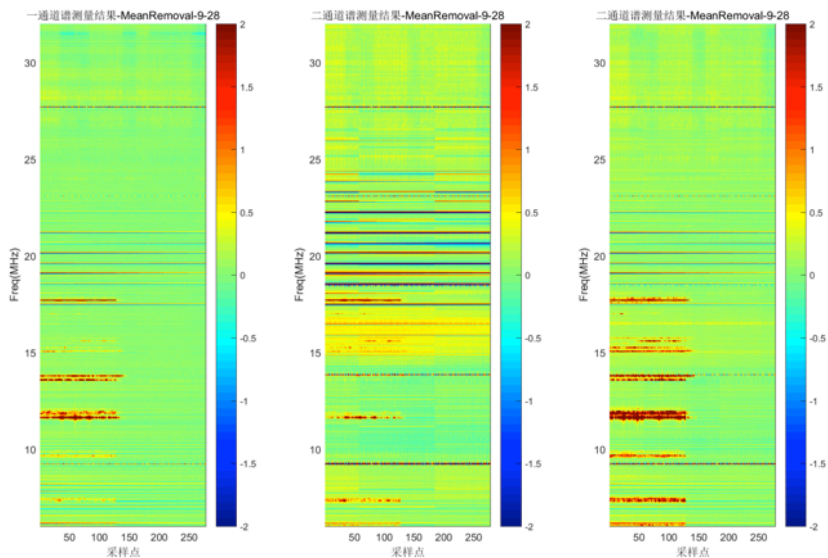

Figure 3-2. The LFI three polarizatio data when the Earth is occulted by the Moon
(provided by NSSC)

Up to Deccember 2018, the total observing time was more than 1000 min, the number of Earth occultation is more than 20, and the number of Jupiter occultation more than 6. Preliminary analysis of the data had showed significant RFIs from the Earth, and the shielding by the Moon is clearly seen. As illustrated in Figure 3-2, three channels of LFI had observed almost identical Earth RFI suppression phenomenon by the Moon.

## 3.2 The CE-4 lander

The far side of the Moon is recognized as the best place for low frequency radio astronomy observations. The Moon can effectively shield radio waves from the Earth, as well as those from the Sun at night. Therefore, low-frequency radio astronomical observation at 10KHz~40MHz offers the opportunity to discover new phenomena and laws in the evolution of celestial bodies. Using the opportunity of Chang'e-4 exploration probe landing on the far side of the moon, a Very Low Frequency Radio Spectrometer is installed on the Chang'e 4 Lander (Figure 3-3). Its scientific mission is mainly to explore the radiating characteristics of the electric fields from radio bursts during the lunar day, and to study the ionospheric characteristics over the landing area.

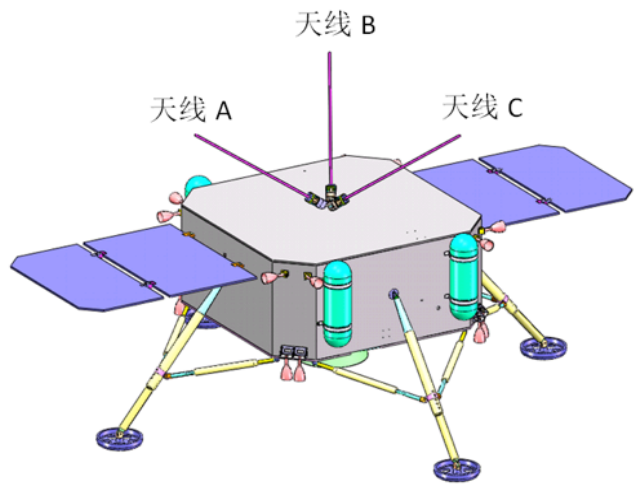

Figure 3-3 Tripole antennas of the very low frequency radio spectrometer
(provided by Institute of Electronics, CAS)

The Very Low-Frequency Radio Spectrometer uses three orthogonal active antennas to receive three components of the electric field of the waves from solar burst or cosmic space. According to the theory of electromagnetic wave propagation, the intensity and polarization characteristics of the total electric field can be obtained by processing of the three electric field components. The frequency spectrum and time-varying information of the electric field can be obtained too. In addition, using the amplitude and phase of the three-component electric field, the direction of arrival of the wave can also be obtained after data processing.

The composition of the very low frequency radio Spectrometer system is shown in Fig. 3-4. As shown in Figure 3-4, the very low-frequency radio Spectrometer is mainly composed of electronic unit, preamplifier, three receiving antennas and cable s. The electronic unit is composed of controller, distributor, clock module, multi-channel receiver, internal calibration module, and interfaces with the Lander. The main performance requirements of low frequency radio spectrum analyzer are shown in Table 3-2.

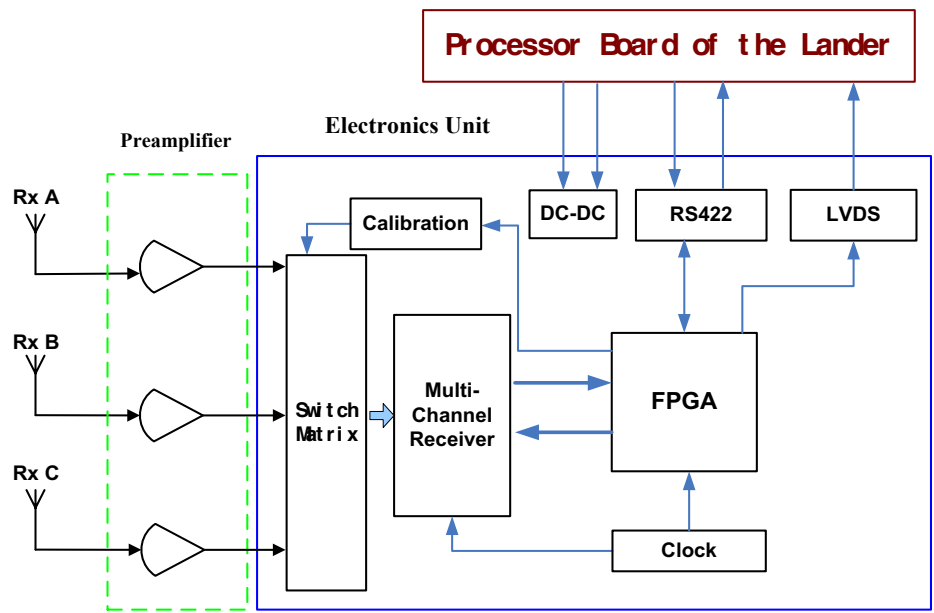

Figure 3-4　System of the very low frequency radio spectrometer (provided by Institute of Electronics, CAS)

On January 3, 2019, Chang'e-4 lunar probe landed on the far side of the moon successfully. At 00:40 on January 4, the receiving antennas A, B and C of VLFRS were deployed smoothly under the control of ground commands, and their deployment lengths were 5 m. At 9:20 on January 5, 2019, the low-frequency radio spectrum analyzer was powered up. The internal calibration and data acquisition modes were carried out respectively. The equipment worked normally, the telemetry data were normal and the scientific data were correct.

Table 3-2 Technical Specifications of VLFRS

| Frequency | 100kHz~40MHz |
|---|---|
| Sensitivity | 7.5nV/√Hz |
| Dynamic Range | ≥ 95dB |
| Frequency Resolution | ≤ 5KHz（100KHz~2.0MHz）<br>≤ 200KHz（1.0MHz~40MHz） |
| Receive Antenna | Three 5m orthogonal monopole antennas |
| Bit-Rate | ≤5Mbps |
| Power | 24W |

## 3.3. The NCLE experiment

The Netherlands-China Low-Frequency Explorer (NCLE) is a low-frequency radio astronomy and space science payload on-board the CE4 relay satellite orbiting the Earth-Moon L2 Lagrangian point. NCLE is a pathfinder mission for a much larger future array of small satellites aimed at observing the earliest phases of our universe, the Dark Ages period in which the first stars and galaxies were formed. Apart from constraining estimates of the global Dark Ages signal, NCLE also aims at observing the Sun and large solar system planets.

The payload is outside the Earth's ionosphere and relatively far away from terrestrial interference although that will still be detectable. NCLE is equipped with three length 5m monopoles that can be configured as (two) dipoles. It covers the 1-30 MHz band virtually inaccessible from Earth for radio astronomers, the band 30-80 MHz allowing cross-calibration with Earth based radio telescopes, and the band above 80 kHz MHz suitable for space science, and planetary and plasma physics. Although the digital electronics is capable to cover the entire band in one go, the band was split in five sub-bands for system linearity reasons (Prinsloo et al. 2018). Currently NCLE is in the commissioning phase.

## 4. The Science Opportunities at Low Frequency

### 4.1 Potential of Discovery

Radio astronomy began in 1933 from the unexpected discovery of a radio noise of extraterrestrial origin detected by Karl Jansky. Throughout its history over more than eight decades, radio astronomy offers numerous examples of unexpected discoveries. These include detections of radio emission from celestial radio sources in continuum (e.g., pulsars, Fast Radio Bursts) and spectral lines (e.g., OH masers). Moreover, almost all major radio astronomy facilities built in the past nearly 60 years demonstrated their superb potential in making discoveries not in the areas for which they had been built. For instance, the Arecibo radio telescope was conceived as a facility for studying back-scattering effects in the ionosphere, and not as a prime astronomical tool for studying pulsars, HI emission and conducting planetary radar experiments. The Very Large Array (VLA) and Westerbork Synthesis Radio Telescope significantly outperformed original expectations as radio imaging instruments in continuum and spectral line regimes. The latest example is the Canadian Hydrogen Intensity Mapping Experiment, CHIME, which has become recently the most productive discoverer of Fast Radio Bursts (FRB's). Even from the name of this experiment it is obvious that the original prime task of the CHIME observatory was unrelated to FRB's. All these and many other examples have in common one major characteristic: each radio astronomy instrument that proved to carry a potential for outstanding, sometimes – transformational discoveries, was built with at least one its main specification significantly exceeding any other instrument. Such the specification might deal with sensitivity, spectral, time or angular resolution, spectrum coverage, etc. In other words, broadening up coverage of at least one major parameter of a new experimental facility practically guarantees outstanding discoveries.

In this sense, there is every reason to expect that a prospective space-based ultra-long wavelength facility will continue the trend of delivering unexpected pioneering discoveries. The reason for such the expectation is simple: just as many other outstanding radio astronomy facilities of the past decades, the ULW radio observatory will open up a large unexplored area of a major parameter of a telescope, its spectrum coverage. The ULW domain is the last unexplored area of the electromagnetic spectrum of cosmic emission. While all declared science tasks of the ULW facility do deserve careful observational studies, unexpected discoveries in the large unchartered spectrum domain have a potential of becoming dominant in the facility's operational agenda. Expect unexpected!

## 4.2 Cosmic Dawn and Dark Ages

The Big Bang left its elusive fingerprints on the cosmos, but the mystery remains. The scientific consensus is that all we see in the visible universe emerged from what is called "inflation", an immensely rapid expansion that occurred some trillionth of a trillionth of a second after the Big Bang. The rapid expansion left a pattern of scattered photons on the microwave sky that have been detected and mapped as millions of tiny ripples, the seeds of all large-scale structure in the universe. However, this approach limits cosmology to at most 0.1 percent precision. This falls well short of what we need to search for the fingerprints of inflation. The only guaranteed signal, primordial non-Gaussianity (PNG), robustly predicted by all inflationary models, requires a thousand fold increase in precision in measuring the PNG parameter $f_{NL}$. The Planck satellite experiment, and any future CMB experiments, are limited at best to $f_{NL} \sim 10$. Future surveys of billions of galaxies will yield $f_{NL} \sim 1$. To improve precision by another 100-fold, we need to go beyond galaxies, to their millions of gas cloud precursors, detectable by searching in the so-called "dark ages" before there were any galaxies, via highly redshifted ($z \sim 50$, the sweet spot for the predicted signal) 21 cm absorption against the CMB. Sensitivity at such low frequencies (~30 MHz) may only be achievable by the far-side lunar radio arrays (Jester & Falcke 2009).

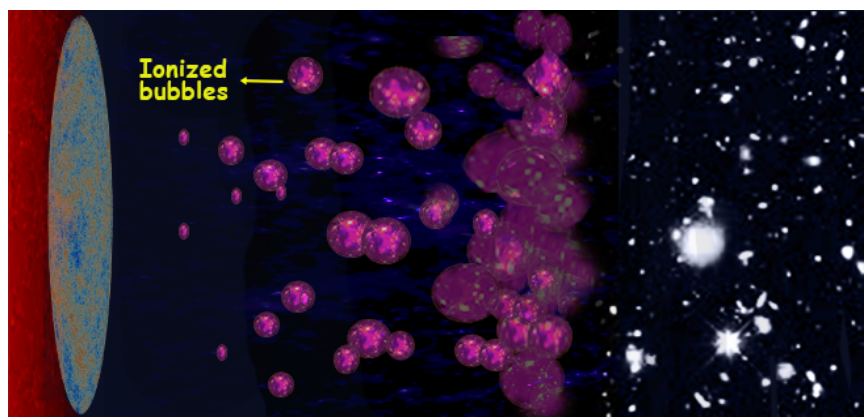

Figure 4-1 The Dark age and Cosmic Dawn (artist impression)

At lower redshift, the shape and the features of the 21-cm signal are tied to the astrophysics of galaxy formation, when the signal during the dark ages (the period prior to star formation) is fully dictated by atomic physics. As soon as the first stars form, their lights dramatically change the environment strongly affecting the 21-cm signal. During cosmic dawn (Figure 4-1), ultraviolet (Lyα) radiation produced by first stars couples the 21-cm signal to the thermal state of the gas, which is cooled by the expansion of the Universe and heated by X-ray radiation of first black holes. In addition, radiation produced by stars at energies above 13.6 eV ionizes neutral hydrogen in the intergalactic medium. These processes are responsible for the deep absorption trough (centered at the frequency of about 60-90 MHz) and the emission peak (at 100-150 MHz) in the global signal (Figure 4-2). Owing to the patchiness of primordial star formation and the finite distance from each source out to which the photons can propagate before being absorbed or scattered, the radiative backgrounds (Lyα, X-ray, ionizing) are not uniform across the sky. These fluctuations result in variability of the 21-cm across the sky. Exotic processes, such as baryon-dark matter scattering (Barkana 2018) or neutrino decay, might affect thermal and ionization histories of the gas, modifying the 21-cm signal from the dark ages, cosmic dawn and reionization. For instance, velocity-dependent scattering cross-section results in enhanced pattern of Baryon Acoustic Oscillations in the 21-cm power spectrum from cosmic dawn (Fialkov 2014). However, in most cases, star formation "contaminates" the signal from cosmic dawn and reionization, and the dark ages remain the optimal period in cosmic history to search for the imprints of the exotic physics.

Due to the large foreground radiation, mapping the fluctuations of the dark age and cosmic dawn requires very high sensitivity and stability, which could only be achieved with a very large area array on the far side of the Moon in the distant future. However, a small array such as the DSL could map the foreground radiation, which would be a useful and necessary first step toward that direction. The signal-to-noise ratio of a global spectrum detection is however independent of the receiver collecting area (for a filling factor of unity, as is the case for three orthogonal dipoles) and hence could be carried out with a modest mission. The global spectrum can be measured to a high precision of interest to cosmology study with a single antenna in a short time, as already demonstrated by experiments such as the EDGES, SCI-HI, SARAS, and so on. The sensitivity however depends on effective control of the systematics. In some aspects, a space-borne experiment can be very advantageous as the problem of ionosphere distortion can be largely avoided. A lunar orbit mission also helps to reduce the effect of ground reflection artifact which may generate fake absorption signals (Bradley et al. 2019). However, the RFIs generated by the satellite itself and any frequency-dependent beam effect must be tamed for the measurement to be successful.

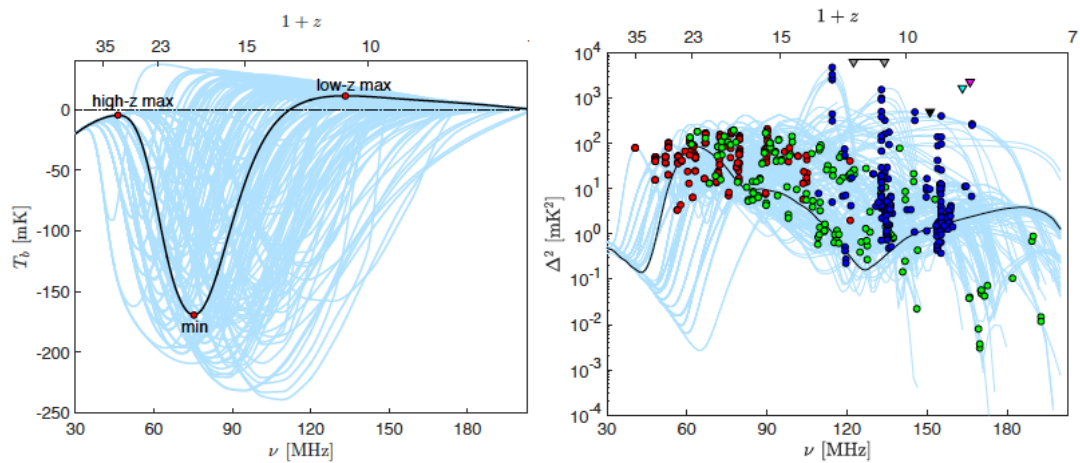

Figure 4-2 The global signal (left) and the corresponding power spectrum (right) for different model parameters. The colored dots on the right panel mark the different stages of evolution: Lyman alpha coupling (red dot), heating transition (green dot) and mid point of reionization (blue dot). The black solid curve is the standard model.
(Cohen et al. 2018)

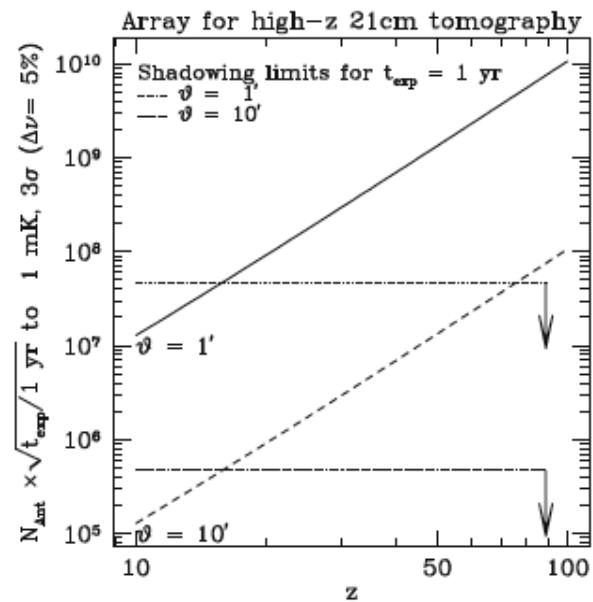

Figure 4-3. The number of cross dipoles required for high-z 21cm tomography (from Jester & Falcke 2009)

## 4.3 Helio Physics and Space Weather

The active Sun exercises a fundamental influence on the Earth's geosystem thereby affecting the quality of life on Earth and the performance of technological systems. Plasma instabilities below the solar surface (chromosphere) sporadically generate bursty releases of energy the most violent of which are identified above the solar surface as coronal mass ejections (CME), clouds of highly ionized plasma ejected into interplanetary space. Despite their great importance to life on Earth, the physical mechanisms governing such events are poorly understood. This is in part due to limited observational capabilities CME in particular to describe 3D structures evolution in time.

Three main types of radio bursts are observed from the Sun particularly in its active state, both related to flares and CMEs. Type II bursts have a frequency drift with time at rates consistent with the speed of the shock through the solar corona and interplanetary medium (~1000-2000 km/s). Type III bursts are emitted by mildly relativistic (~0.1 - 0.3 c) electron beams propagating through the corona and interplanetary space that excite plasma waves at the local plasma frequency. Their frequency drift rate is much higher than that of Type II bursts. Type IV bursts are emitted by energetic electrons in the coronal magnetic field structure.

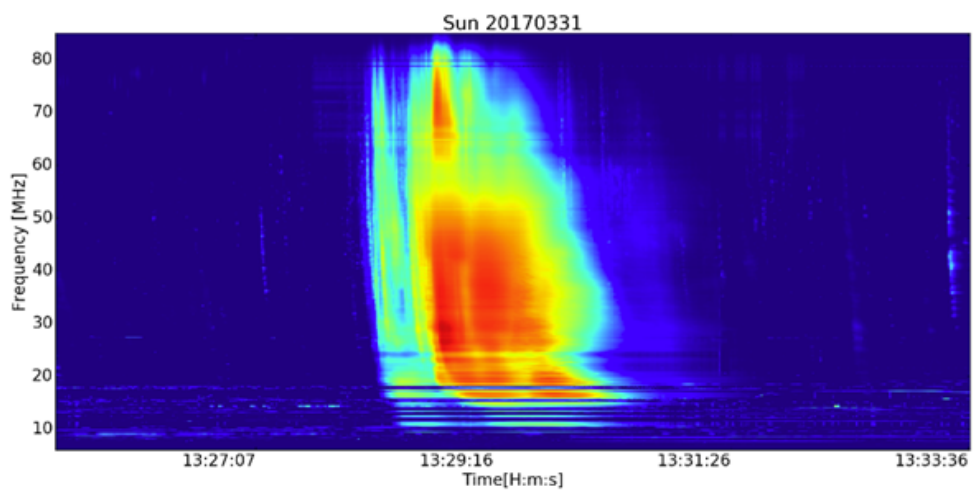

Figure 4-4. An example of solar burst detected by PL610 single LOFAR station P (figure provided by Space Research Center, Polish Academy of Sciences)

To develop a quantitative model of energy transfer from Sun to the ionosphere-magnetosphere system it is necessary to consider the plasma wave interaction. It is very hard to judge which physical process dominates during a geomagnetic disturbance, in particular it seems necessary to bind the observations from different region of Sun-Earth system.

The magnetosphere-ionosphere-thermosphere subsystem is strongly coupled via the electric field, particle precipitation, heat flows and small scale interaction. The wave-particle interactions in radiation belts region are one of the key parameters in understanding the global physical processes which govern the near-Earth environment. The efficiency of the solar driver depends on the prevailing specific properties and preconditioning of the near-Earth environment.

By providing dynamic spectra and detailed imaging of the solar radio emissions, future missions such as DSL will allow monitoring and modelling of plasma instabilities in the solar corona and wave-particle interactions in the activity centres of the Sun. DSL will offer great opportunities for radio studies of the solar wind and the heliosphere. It will permit observations of solar radio bursts at low frequencies with much higher spatial resolution than possible from any current space mission. It will also allow observations much further out from the solar surface than possible from the ground, where the ionosphere confines the field of view to within a few solar radii. DSL will dynamically image the evolution of CME structures as they propagate out into interplanetary space and potentially impacts on the Earth's magnetosphere.

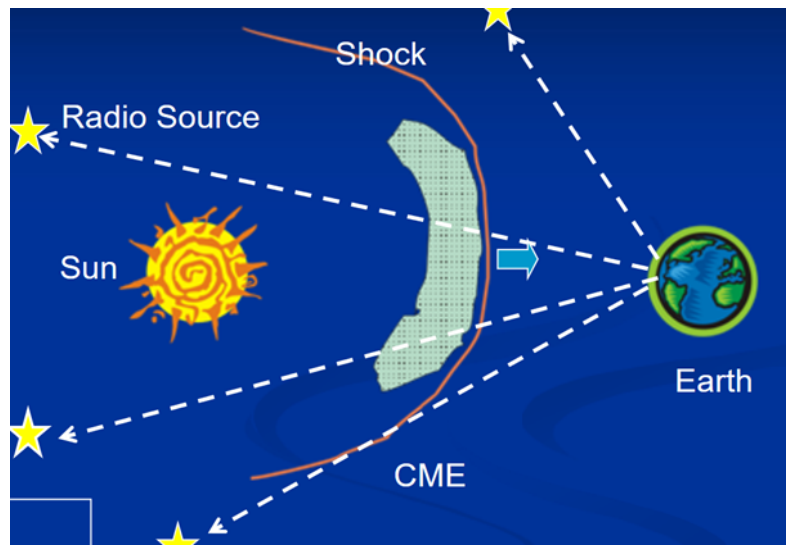

Figure 4-5 The IPS measurement (figure provided by Space Research Center, Polish Academy of Sciences).

In combination with interstellar plasma scintillation (IPS) measurements from terrestrial radio telescopes, particularly LOFAR, it should be possible to distinguish between the relative contributions of shocks and ejecta in interplanetary CMEs and track the evolution of these structures separately. The combined *in situ* and remote solar observations could in the next decade give us a much better understanding of heliophysics, solar-terrestrial physics, and space weather. DSL will also provide an ideal low-noise facility for IPS observations in its own right, with the additional virtue of providing heliospheric calibration for all-sky astronomical measurements.

Complementary terrestrial LOFAR observations will provide the heliospheric context for DSL diagnostics. The ongoing project LOFAR4SW in the frame of European H2020 program will deliver the full conceptual and technical design for creating a new leading-edge European research facility for space weather science. In a major innovation, LOFAR4SW will prepare for a large scale high-end research facility in which completely simultaneous, independent observing modes and signal paths provide continuous access to two research communities: radio astronomy and space weather research.

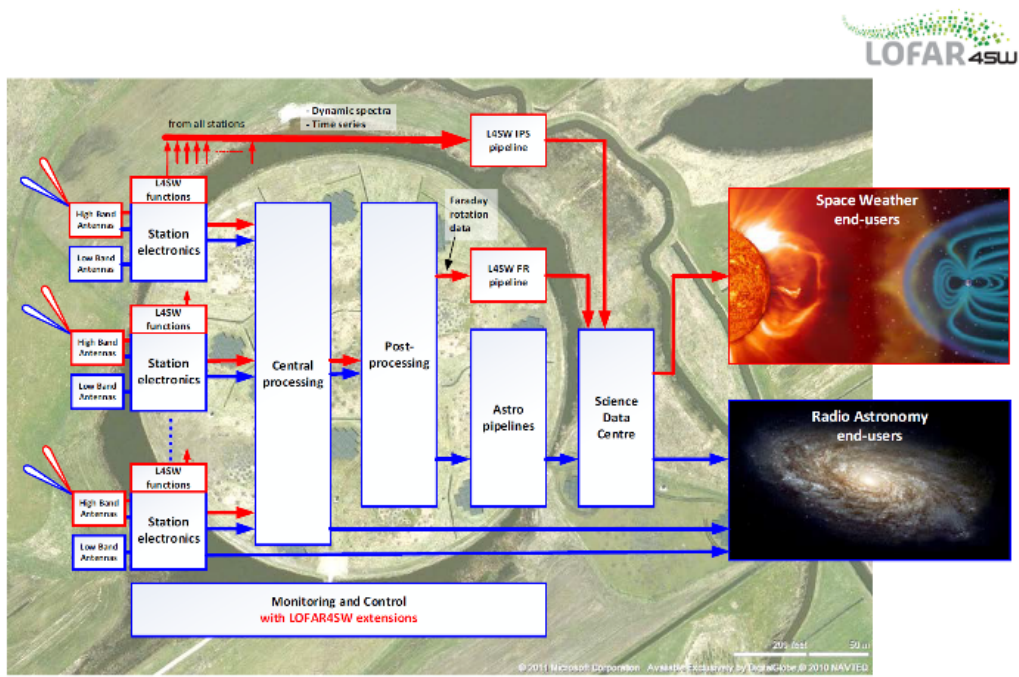

Figure 4-6 The LOFAR4SW experiment (figure provided by Space Research Center, Polish Academy of Sciences)

## 4.4 The Planets, Exoplanets

The Earth and the four giant planets in the solar system have magnetospheres, where energized keV-MeV electrons produce intense non-thermal low frequency radio emissions in the auroral regions near and above the magnetic poles, and these affected by the solar wind and satellite interactions. The solar system radio sources are very bright, Jupiter and Sun have equivalent brightness in decametric range. The planetary radio emissions are intense but sporadic and strongly anisotropic. The Jupiter is alway active, and emits up to 40 MHz which can be observed from the ground. The other planetary radio sources are emitting primarily below 1 MHz. which puts them far below the cutoff frequency of the Earth ionosphere. The Cyclotron Maser instability (CMI), operating at the local electron cyclotron frequency in the magnetospheres of these radio-planets, is confirmed as the general mechanism for radio emission at the polar regions. Planetary lightning is another source of low frequency radio emission.

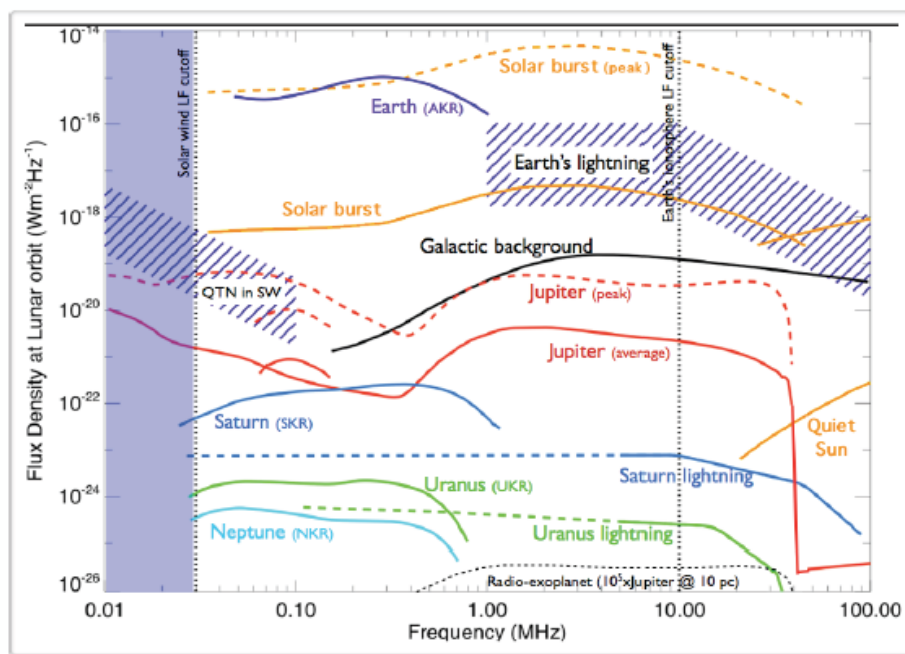

Figure 4-7 The flux density of various low frequency sources

The radio waves may also be produced by exoplanets via similar mechanisms. If detected, they would provide many parameters on the exoplanetary system. The observation would however be very difficult, many spacecraft and integration will be required.

Ground and space measurement are complementary. In particular, space-borne observation and monitoring may provide useful information at the lower frequencies. The fine structure of the radio emission provides remote sensing capability for auroral plasma study. The DSL is a near-linear array and has limited snapshot-imaging capability, nevertheless it can observe the planetary emission on hourly time scale. In future, more systematic lunar orbit observation may also be achieved with a swarm of satellites.

**4.5 Cosmic Ray and Neutrinos**

The research on nature and sources for the highest energy particles is one of the major topics of modern high-energy astrophysics and particle physics. To date a couple of tens of cosmic ray events have been observed with energies in excess of 1020 eV, these are the so-called ultra-high-energy cosmic rays (UHECR). According to our current understanding of the Universe, such particles should not exist because of the spectrum cut-off $\sim$ 6 × 1019eV caused by the Greisen-Zatsepin-Kuzmin (GZK) effect. In any case, these particles need further experimental investigation. Detection of UHE neutrinos is also of great importance for both astrophysics and particle physics. Since the neutrinos are chargeless weakly interacting, ultrahigh energy neutrinos (UHEN) can propagate unaffected over cosmic distances, and therefore their arrival directions carry direct information on their sources. Observations of UHEN would open up a new window on the highest-energy astrophysical process.

The big problems of both UHECR and UHEN event detection are not from the character of the event, but from their extreme rarity. The Moon offers a very huge natural detector volume, and is first proposed by G.A. Askaryan (Askaryan 1962) to be a target to detect showers initiated by cosmic ray and neutrinos. When a high-energy particle interaction occurs in a dense medium like ice, rock salt, lunar regolith, and the atmosphere, it draw electrons off the surrounding medium and transferring them into the shower disk. With the annihilation of shower positrons in flight, there will be a net excess of electrons. These fast excess electrons can emit radio waves through the Cherenkov mechanism. Using this mechanism many experiments have been performed or proposed to detect high-energy cosmic particles, as GLUE (Gorham et al. 2004), LUNASKA (James et al. 2010), and NuMoon (Scholten et al., 2009) etc., which are all based on the terrestrial telescopes. However, the radio signal attenuation over the long distance between the Moon and the Earth severely limits the detections, and the refracting of radio waves by the ionosphere will lose partially its original location information as well. Compared with the terrestrial radio telescopes, a lunar orbit radio telescope will be more competitive since it is close to the lunar surface and lack of the atmosphere. Preliminary studies show that a DSL-like array operating at the frequencies (<30 MHz) with a 300 km lunar orbit can detect dozens of UHECR events and few UHEN events per year. With the decreasing of the orbit height, the UHECR detections will be increased greatly. The joint observations between multi elements of the array will further provide the possibility to retrieve the UHECR or UHEN origins.

### 4.6 Pulsar and ISM

Pulsars are among the best-studied cosmic radio sources. They are, however, intrinsically dim below 30 MHz, and their impulsive signals are strongly affected by interstellar dispersion and scattering, perhaps only a few brightest ones are detectable with a mission such as DSL. Detection of these sources will allow for groundbreaking research into low-frequency pulsar emission. Because of the atmospheric cutoff at ~10 MHz, only space-based observations can shed light on this low-frequency behavior. These would be a valuable complement to the detailed work currently produced by LOFAR. The data collected in the course of the DSL mission life can produce folded, coherently de-dispersed profiles for the dozen or so brightest nearby pulsars in the sky.

Such observations will also provide insight into the scattering properties of the ISM at low frequencies. The interstellar medium (ISM) is the matter that exists in the space between the star systems in the Galaxy, including gas in ionic, atomic and molecular form, as well as dust and cosmic rays. The ISM exists in multiple phases, with different temperature and ionization state. It provides the material from which the stars and planets form, and it is also constantly replenished with newly accreted gas as well as material and energy from stellar winds, planetary nebulae and supernovae. The ISM plays a crucial role in astrophysics. However, diffuse ionized gas in particular has been very difficult to observe until now. The very low frequency observations of DSL will provide an excellent means of tracing the ISM in this phase. Besides information dervied from the pulsar dispersion measure, the new multi-frequency all-sky map at the ultralong wavelength where ISM absorption become significant and highly sensitive to frequency will reveal the distribution of the ISM, showing the ionized gas around the solar system in detail.

### 4.7 Extragalactic Radio Sources

One of the major goals of an ultra-low frequency survey would be building the first catalog of extragalactic radio sources at frequencies lower than 30 MHz. Most sources in the radio sky are extragalactic sources, mainly including active galactic nuclei, radio galaxies, and galaxy clusters. However, there is till barely any observational constraints on the spectral energy distribution of them at frequencies lower than 20 MHz. Any observations of the spectral indices of these sources will constrain the mechanisms of radiation and absorption, and shed light on the nature and environments of the sources. In deed, the typical synchrotron spectrum would change to synchrotron self-absorption spectrum if the source were sufficiently compact and optically thick (Jester & Falcke 2009). The strength of magnetic field and the free-free absorption by the local electrons can also modify the observed spectral energy distribution. In addition, due to the spectral aging effect (Alexander and Leahy, 1987; Blundell and Rawlings, 2001), lower-frequency observations probe the older parts of a source and can therefore be used to constrain the age of a source.

Galaxy clusters are extended radio sources, where the plasma in the intracluster medium emits synchrotron radiation. However, the particle acceleration mechanisms in galaxy clusters are unknown. Low frequency observations could probe the giant radio halos in merging clusters, or mini halos in cool-core clusters. The steep spectra of cluster halos imply that lower frequencies are more promising to detect the cluster halos, specifically for low mass and high-redshift objects (Cassano et al. 2006, 2008). Savini et al. (2018) have found the first evidence of diffuse ultra-steep-spectrum radio emission surrounding the cool core of a cluster, which means that under particular circumstances, both a mini and giant halo could co-exist in a single cluster. Observational constraints on the spectral properties and the sizes of them could further constrain the formation history and the magnetic fields of clusters.

**4.8 SETI**

The Search for Extra-terrestrial Intelligence (SETI) is always important for humans' curiosity, and aims to test the hypothesis that extraterrestrial civilizations emit detectable signals using a technique similar to what we have, preferentially in radio band (e.g. Croft et al. 2018). A major issue with the detection of SETI signals is to distinguish the signal from terrestrial radio frequency interference (RFI) and other astrophysical signals. Using the Moon to shield the RFI from the Earth, the DSL project could significantly reduce the number of false positives from terrestrial transmissions, and increase the confidence in the detection. The advantages of using an interferometer array with a wide field of view in searching for a SETI signal are highlighted in Garrett (2018). Although the requirements of high time and frequency resolution can be quite a challenge with the current technologies, the DSL array observing at the far side of the Moon will be a very promising complementary to the bands covered by the ground-based experiments, placing extra constraints on the prevalence of civilizations in the Universe.

## 5. The DSL mission

The Discoverying Sky at the Longest wavelength (DSL) concept consists of a constellation of satellites circling the Moon on nearly-indentical orbit, forming a linear array while making interferometric observations of the sky. A major goal of the mission is to map the sky below 30 MHz using the constellation of satellites as an interferometer array. Another goal of the mission is to make high precision global spectrum measurement over the frequency range of interest to study of the dark age and cosmic dawn, as it is not affect by the ionosphere disturbance and RFIs in ground observation. The linear configuration allows the relative positions of each satellite to be measured with very limited instrumentation--star sensor camera for angular measurement, and microwave line for distance measurement, the latter also serves as inter-satellite data communication and synchronization system. The constellation includes one mother satellite and a number of (tentatively set as 8) daughter satellites. The daughter satellites are each equipped with electrically short antennas and receivers to make interferometric observations, while the mother satellite will collect the digital signals from the daughter satellites for interferometry correlation, and transmit the data back to Earth. It also has the high frequency band spectrometer which shall make the global spectrum measurement in the frequency above 30 MHz. At present we consider a circular orbit of 300 km height which is sufficiently stable against lunar gravity perturbations.

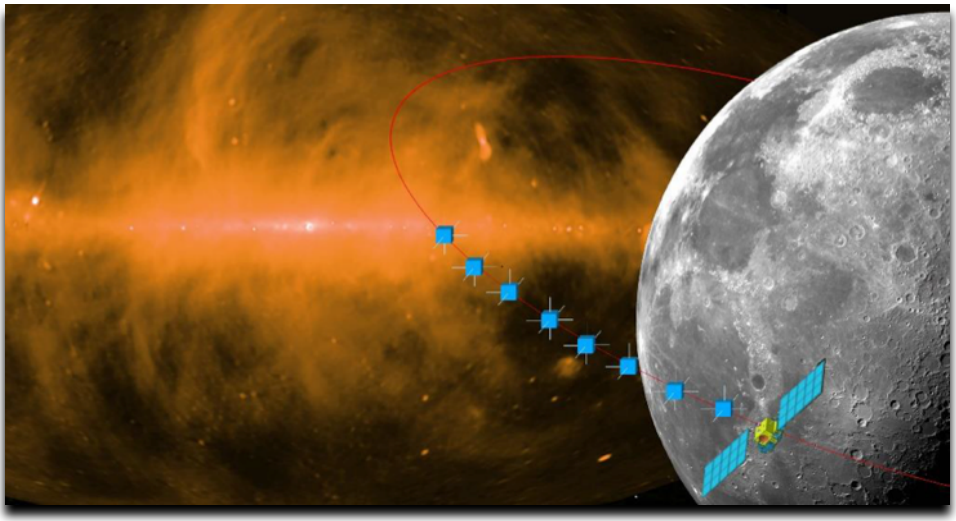

Figure 5-1 An artist's concept of the DSL mission

## 5.1 Array Configuration

The constellation is designed to be reconfigurable in orbit, allowing interferometric observations with a range of different baselines formed between the satellites. The daughter satellites will be docked in the mother satellite during launch and lunar transfer, then sequentially released after entering into the lunar orbit to form the linear array. An artist impression is shown in Figure 5-1.

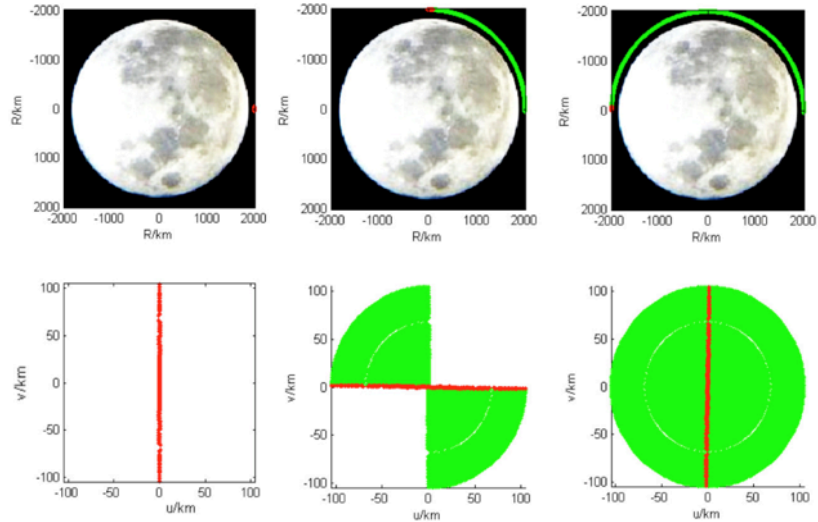

Figure 5-2 Evolution of the spatial aperture plane filling after one orbit of DSL (down) as it orbits the Moon (up)

The daughter satellites form multiple baselines, Figure 5-2 shows the evolution of the baseline vectors between the daughter satellites as the array circles around, generating concentric rings in the so called uv plane (u, v are Cartesian coordinates in units of wavelength). The gaps between the rings can be partially filled by using bandwidth synthesis and by varying the relative distance between the satellites. It should be emphasized that the satellites do not need to have fixed relative positions, as long as the positions can be determined the array could work.

As the short dipole antenna are sensitive more or less in all directions, and the Moon only shields a small fraction of the sky, there is a mirror symmetry between the two sides of the plane which can not be distinguished using the interferometry data of a single oribit alone. However, the orbital plane precesses a full 360 degrees in 1.29 years, so after a few orbits, the aperture plane will be tilted. Sources in both halves of the hemispheres will have different projections on the two aperture planes and can in principle be separated. Figure 5-3 shows the cumulative filling over time (many orbits) of the aperture with visibility sample points. After 360 degree precession, the visibility filled a 3D "doughnut" shaped structure. This dataset is similar to a 3D hologram, and a sky image can be reconstructed by linear invertion（Huang et al. 2018), as shown in Figure 5-4.

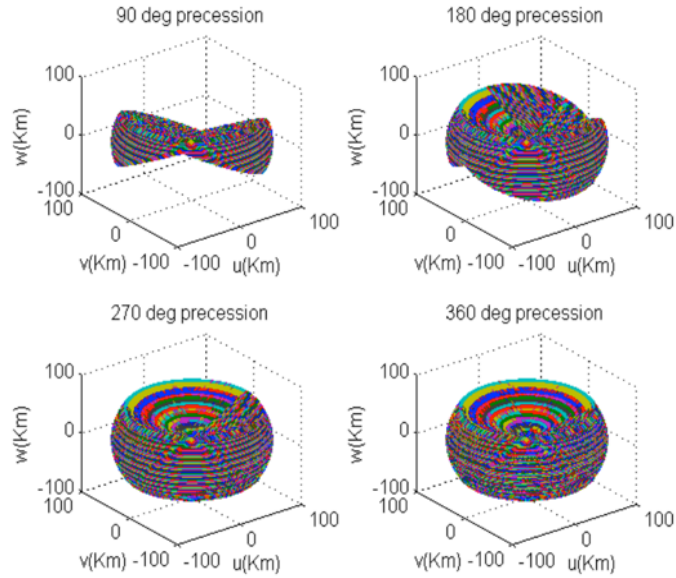

Figure 5-3 Cumulative filling of the 3D aperture over respectively 90, 180, 270 and 360 degrees of orbit precession

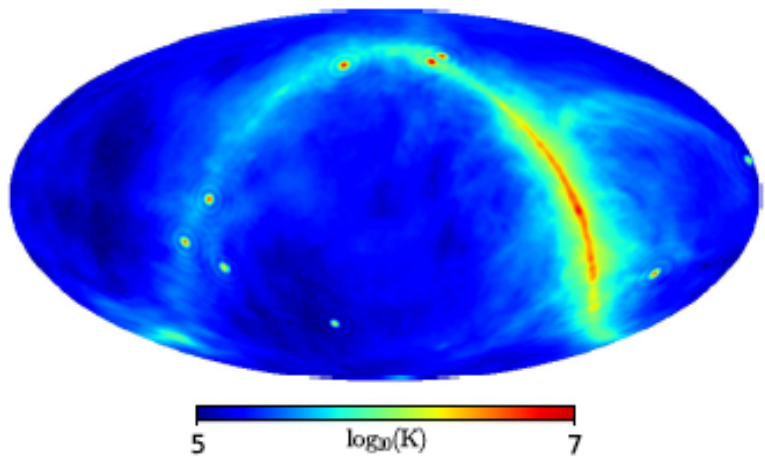

Figure 5-4 A Reconstructured sky map with simulated DSL observation. A few bright spots are added to the lower left part of the map the to illustrate the reconstruction for point sources. (Q. Huang et al. 2018)

The angular resolution of the array is determined by the maximum baseline. However, interstellar medium (ISM) scattering can broaden the point source to about 0.5 degree at such low frequency, and the broadening by interplanetary medium (IPM) may be even larger (Jester & Falcke 2009). The maximum baseline for DSL is set as 100 km, which at 1MHz corresponds to a resultion of 0.17 degree. Further increase of the maximum baseline would not improve the imaging resolution, but the technological difficulty in position measurement and data communication would increase significantly. Most of the known radiation mechanisms at low frequency such as the synchrotron and free-free radiation produce continuum radiation, for which high spectral resolution is not needed. The foreground subtraction of the redshifted 21cm line also only requires a moderate spectral resolution. An exception to this is found in the radio recombination lines, which requires a sub-kHz spectral resolution. However, as the satellite array is circulating the moon and the baselines are moving, fine spectral resolutions are required to keep coherence.

Detection sensitivity at long wavelengths is strongly influenced by the confusion of multiple sources nominally above the sensitivity. Confusion limits identification and measurement of individual sources, and occurs when there is more than one source in every 30 synthesized beam areas. Extrapolation of source count statistics of the VLA 74 MHz and the Parkes 80/178 MHz Catalog lead to an estimated one sigma confusion noise limit of 5.6 Jy and 65 mJy for 1 MHz (10' beam size) and 10 MHz (1' beam), respectively. After observing with 8 antennas at 1 MHz (1 MHz bandwidth) for a few weeks, DSL will reach the confusion noise limit above, which effectively sets the limit to imaging sensitivity of DSL. At 10 MHz mapping sensitivity is limited by integration time primarily, not by confusion.

For the Dark Ages science case, the goal is to reach 5 mK error on the brightness temperature estimate at 20 MHz, within eight months of mission duration: this would allow a detection of the dark age signal at SNR=10. Reaching this 5 mK accuracy requires >60 dB pass-band calibration. For foreground subtraction, the full band needs to be observed with a spectral resolution of 1 MHz. We expect that these need to be combined with higher-frequency measurement of the global sky from the ground and possible from space (i.e. DARE at >40 MHz) to cover the full 20-100 MHz range.

Table 5-1. Summary of DSL requirements

| No | Requirement | Value |
|---|---|---|
| 1 | Frequency range | 1~30MHz for imaging<br>30~120MHz for high precision spectrum |
| 2 | Spectral resolution | 1MHz for imaging<br>10kHz for high precision spectrum |
| 3 | Total integration time | 1.29y for imaging |
| 4 | Baselines | Maximum length 100km<br>Minimum length 100m |
| 5 | Mission lifetime | 5y |

A summary of the DSL requirements is listed in Table 5-1. It basically covers all requirements the science cases described above, appended with a few requirements stemming from technical considerations as will be explained in the following sections.

## 5.2 Satellite and Payloads

The mother and daughter satellites are launched together to lunar orbit. As illustrated in Figure 5-5, the upper part of the mother satellite has deployable solar panels, high gain antenna (for ground communication), high gain ISL antenna and other sensors and antennas, and on top of these sit the cone-shaped high frequency spectrometer (HFS) antenna and its ground plane which shield it from the rest of the mother satellite. The lower part of the mother satellite carries the ring of daughter satellites on its external sides. This combo sits above a propulsion module, which propels the satellites during the Earth to Moon transfer and separated after entering the lunar orbit. The daughter satellites are then released one by one.

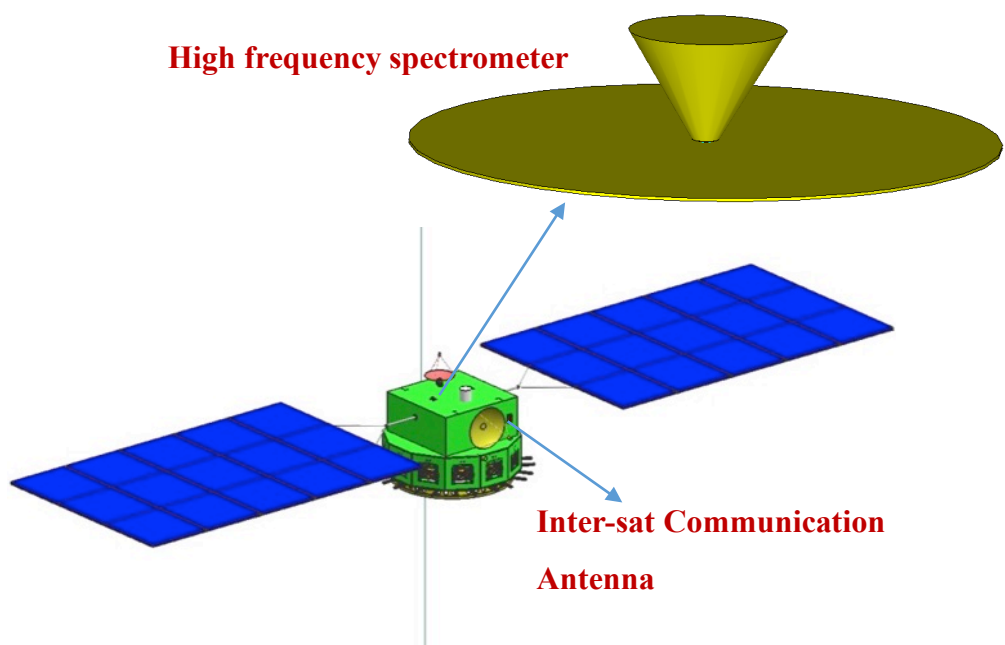

Figure 5-5 A split diagram of he Mother satellite

To measure the 21cm global spectrum in the frequency range 30-120MHz, a dedicated spectrometer is placed on the mother satellite. A frequency-independent beam is the first consideration for antenna design, and thermal effect and mechanical strength will also be taken into account. A deployable ground plate will be mounted right under the antenna to ensure the beam pattern not affected by the satellite itself. A precise calibration system will be embedded into the receiver as a core module, and differential measurement will be used in the receiving system.

Each daughter satellite is an identical cuboid, which carries a low frequency interferometer and spectrometer(LFIS), as shown in Figure 5-7. The daughter satellites have three axis stabilized with respect to the Moon, with pointing accuracy better than 1°, attitude stability better than 0.1°/s, and attitude measurement better than 0.01°. Each has three orthogonal short dipole antenna, a receiver, and a digitizer. Limited by the inter-satellite communication bandwidth, a selection of 1000 narrow bands (1 kHz) within the 1-30 MHz range are used for interferometry.

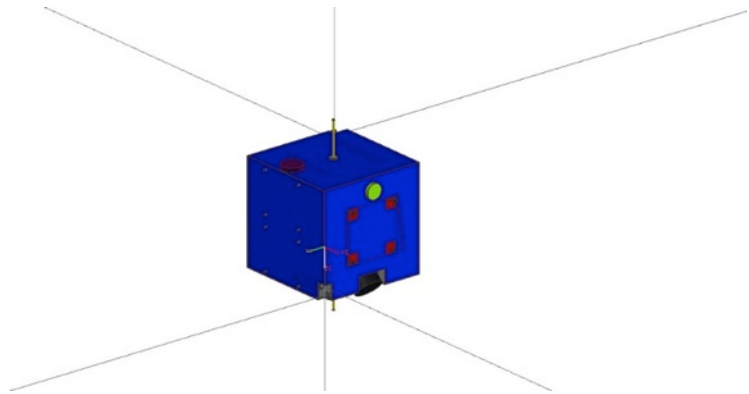

Figure 5-7 Daughter satellite with LFIS antenna deployed

Table 5-2. Summary of LFIS parameter

| No | Item | Value |
| --- | --- | --- |
| 1 | Antenna | 2.5m three-dipoles |
| 2 | Frequency range | 1MHz-30MHz |
| 3 | Spectral resolution points | 8192 |
| 4 | Receiver gain stability | 0.02dB |
| 5 | Sensitivity | 1000K@30MHz（1MHz，10minutes integrated time） |

The satellites will naturally have some velocity differences under the irregular gravity field of the Moon, and without control they may run away. The formation is automatically controlled to keep the linear array stable. A reconfiguration strategy which balances the fuel consumption among all daughter satellites include two steps, step 1: keep S1 stationary, S2-S8 shrink; step2: keep S8 station, S1-S7 expand, as shown in Figure 5-8. We estimate the reconfiguration period is about 20 days, each satellites needs 80mm/s delta-V in each formation reconfiguration control period.

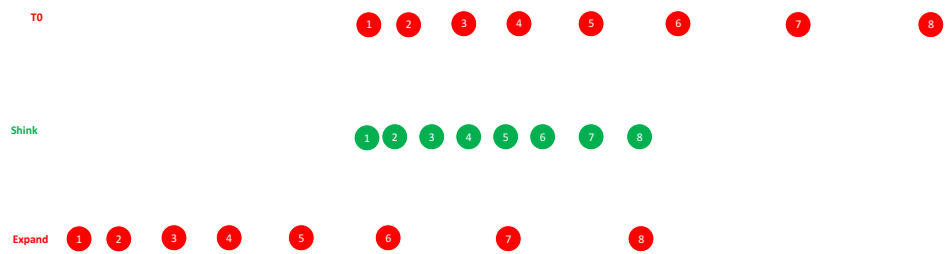

Figure 5-8 Formation reconfiguration strategy

The inter-satellite communication, ranging, clock synchronization and angular measurement is provided by the inter-satellite dynamic baseline apparatus (ISDBA). It uses microwave link (Ka band) for data communication (40 MHz) between the mother and daughter satellites. The same microwave line is also used for ranging and synchronization by using the dual one-way ranging (DOWR) principle (Figure 5-9). The scope of inter-satellite baseline is from 100m to 100km in order to meet the above-mentioned science requirements. For the angular measurement, an LED light array is put on the mother satellite for identification, and the star sensor cameras on the daughter satellites will take photos in the direction of the mother satellite against background stars, the baseline direction is determined by comparing with star map (Figure 5-10). The baseline positioning precision is set to be 1m each direction (1/10 wavelength at 30 MHz). This requires a ranging precision of 1 m and an angular precision of 10urad at 100km. The parameters are summarized in Table 5-3.

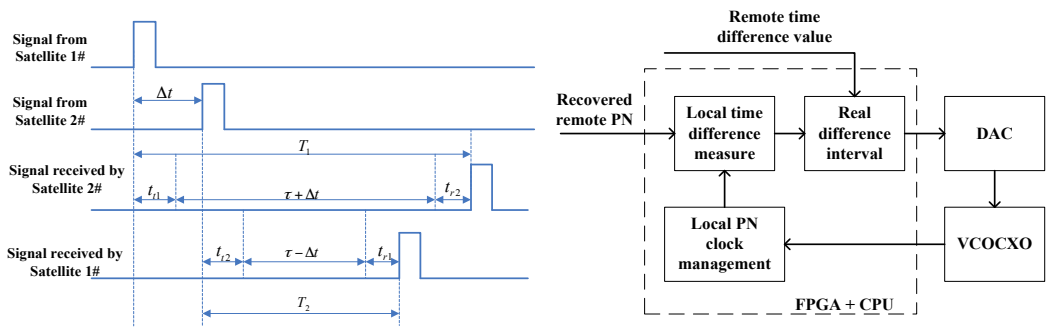

Figure 5-9. The principle diagram of dual one way ranging measurement

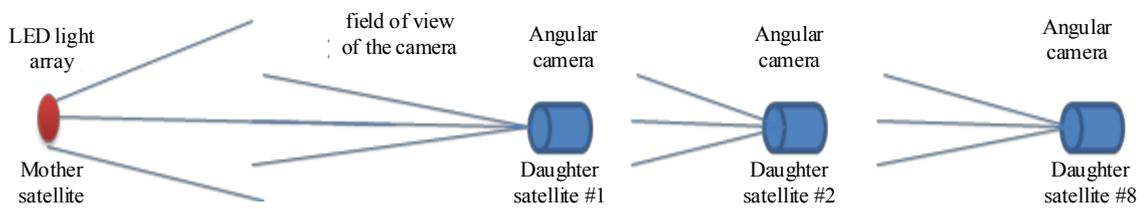

Figure 5-10 Angular position determination

Table 5-3 ISDBA parameters

| No | Item | Value |
|---|---|---|
| 1 | Communication distance | 1km-100km |
| 2 | Ranging accuracy | 1m@100km |
| 3 | Clock synchronization | 3.3ns@100km |
| 4 | Inter-satellite data transmission rate | Adjustable, up to 40Mbps for each daughter satellite |
| 5 | Angular accuracy | 10urad@100km |

## 6. Synergies

A number of both ground-based and space-borne low frequency radio experiments are currently operating or being planned. These experiments are aimed for many different science objectives, from the detailed study of the Sun and planets and space weather, to the exploration of the dark ages and cosmic dawn, and adopted different approaches in their design, so that they provide both independent checks and also complementary to each other.

### 6.1 Ground Global Spectrum Experiments

Examples of current ground-based Global 21-cm experiments operating between 50 and 200 MHz include: the Experiment to Detect the Global Epoch of Reionization (EoR) Signature (EDGES, Bowman et al. 2018a), the Shaped Antenna measurement of the background RAdio Spectrum (SARAS, Singh et al. 2018), the Sonda Cosmológica de las Islas para la Detección de Hidrógeno Neutro (SCI-HI, Voytek et al. 2014), the Large-aperture Experiment to detect the Dark Ages (LEDA, Price et al. 2018), Probing Radio Intensity at high-z from Marion (PRIzM, Philip et al. 2019), and the Cosmic Twilight Polarimeter (CTP, Nhan et al. 2019).   Monsalve et al. (2019) summarized the results from these single-antenna experiments and several radio arrays, all of which are upper limits with the exception of EDGES.   These include:

- An upper limit for the absorption amplitude of 0.89 K at 95% confidence by LEDA (Bernardi et al. 2016);
- Upper limits of 1-10 K by SCI-HI (Voytek et al. 2014);
- Upper limits on the power spectrum by MWA (Ewall-Wice et al. 2016) and by LOFAR (Gehlot et al. 2018).

#### 6.1.1. The EDGES experiment

The Experiment to Detect the Global EoR Signature (EDGES) is a pioneering experiment that has measured the sky-averaged radio spectrum since 2006 with the objective of detecting the predicted global 21-cm signal from the cosmic dawn and the epoch of reionization (EoR). It observes from the Murchison Radio-astronomy Observatory (MRO) in Western Australia, which is one of the most radio-quiet sites on Earth. Due to the daunting challenge resulting from the small amplitude of the 21-cm signal compared to the strong diffuse Galactic and Extragalactic foreground, EDGES has steadily focused on improving the instrument calibration until it reaches extremely high accuracy. As a result of these efforts, in 2010 EDGES placed the first 21-cm constraints on the duration of the EoR, disfavoring a sharp and sudden decrease of the average hydrogen neutral ([Bowman & Rogers 2010](#)). Since 2015, using more mature and sophisticated instruments EDGES has measured and characterized the astrophysical foregrounds, reported effects from the ionosphere, and constrained millions of phenomenological and physical models for the 21-cm signal.

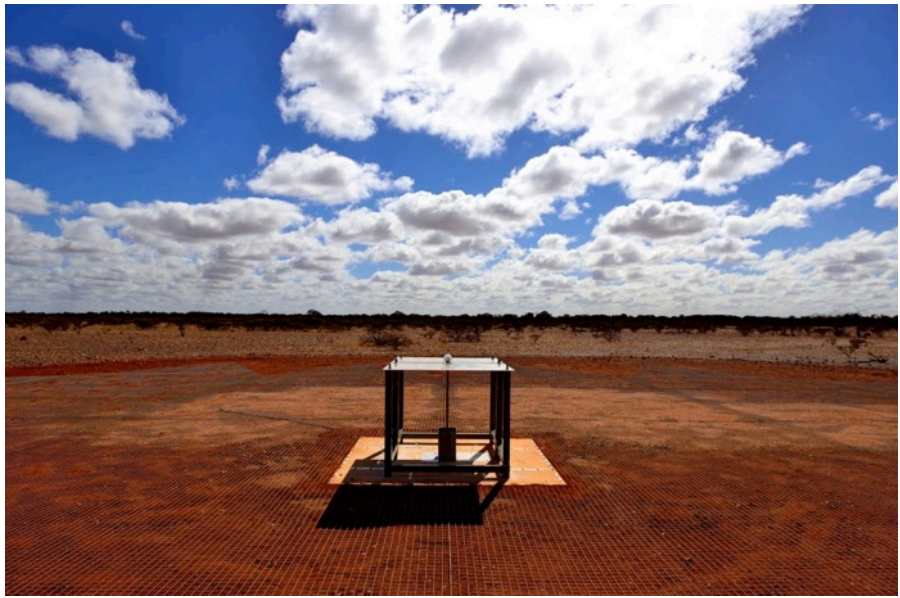

Figure 6-1 An EDGES antenna.

In 2018, using two instruments measuring in the range 50-100 MHz EDGES reported an absorption feature in the spectrum with an amplitude of about 500 mK, centered at 78 MHz, with a width of 19 MHz. This feature is significantly deeper than all theoretical predictions, and has a flattened bottom. It was also reported to be relatively stable over 24 hours of sidereal time, which is an important requirement for the global 21-cm signal. If this feature is interpreted as the 21-cm signal, it would be centered at a redshift z~17; it would also reflect that significant Lyman-alpha coupling of the 21-cm spin temperature to the kinetic temperature of the intergalactic medium (IGM) was occurring approximately 180 Myrs after the Big Bang. This would represent the detection of the effect from the first generations of stars in the Universe during cosmic dawn.

Independent confirmation of this detection by different experiments, and especially from space, would open a new window on the early Universe that will allow us to shed light on an exciting and unexplored period of its evolution. The deep absorption has already motivated hundreds of theorists to propose new physical processes in the early Universe, most of which produce a colder IGM or a stronger radio background than previously thought. Due to the interest resulting from this detection, the EDGES measurement has been recognized by the Physics World magazine as one of the Top 10 Breakthroughs in Physics in 2018. Since the publication of the detection paper, EDGES has continued measuring from the MRO with different antennas and conducting calibration cross-checks. All these recent measurements provide additional evidence for the absorption feature. EDGES is working to finalize the analyses and report these new tests and observations.

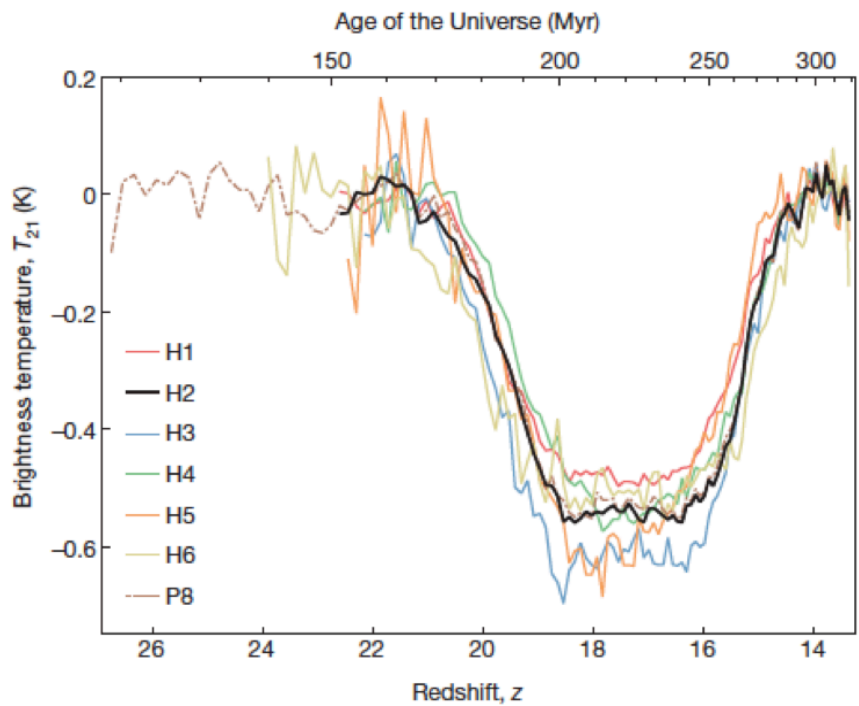

Figure 6-2. The EDGES absorption spectrum (J. Bowman et al. 2018a).

Ground-based experiments are challenging because of unaccounted effects arising from difficult observations within the environment and with the low frequency radio instrument. For example, ionospheric effects (e.g., Datta et al. 2016; Vedantham & Koopmans 2015) cause absorption, refraction, and re-emission which grow dramatically below 50 MHz. Radio frequency interference on Earth is a concern. RFI is scattered/reflected around the globe by the ionosphere as well as by spacecraft debris in LEO. Even human-made RFI reflections off the Moon are seen by ground-based radio telescopes at remote locations (Burns et al. 2017).

Importantly, antenna beam effects due to complex interaction between the ground and the foreground (e.g., Mozdzen et al. 2018; Bradley et al. 2019) are a major problem with ground-based experiments. Based upon their modelling of the publicly-available EDGES data, Hills et al. (2018) claimed that the 78 MHz trough reported in Bowman et al. (2018a) is not unique and implies unphysical fitted parameters for the ionosphere and astrophysical foreground. In their reply, Bowman et al. (2018b) argue that degeneracies between fitted parameters and systematics could explain some of the (Hills et al. 2018) effects. In addition, Bradley et al. (2019) found that small imperfections in ground screens can lead to systematic artifacts that mimic broad 21-cm features in the spectrum. Such a ground screen is not needed for space-based missions.

### 6.1.2. SARAS

An attempt to use Shaped Antennas to measure the background RAdio Spectrum – the SARAS experiment – evolved from radiometers with fat-dipole antennas (Patra et al. 2013) to monopole antennas of different geometries (Singh et al., 2018b), shaped for wideband spectral measurements with maximally smooth (Satyanarayana Rao et al. 2017) reflection efficiencies. These were deployed in radio quiet sites in India, at the Timbaktu Collective in Southern India and in the Tibetan Plateau in Ladakh.

SARAS measurements of the radio sky spectrum in the 110-200 MHz band were analyzed looking for evidence for any of the redshifted 21-cm signals from Cosmic Dawn and Reionization theoretically predicted by Cohen et al. (Cohen et al. 2017). The class of cosmological models in which heating of primordial gas is inefficient---leading to deep absorption signals---together with rapid reionization was rejected by the SARAS measurements (Singh et al., 2017，2018a).

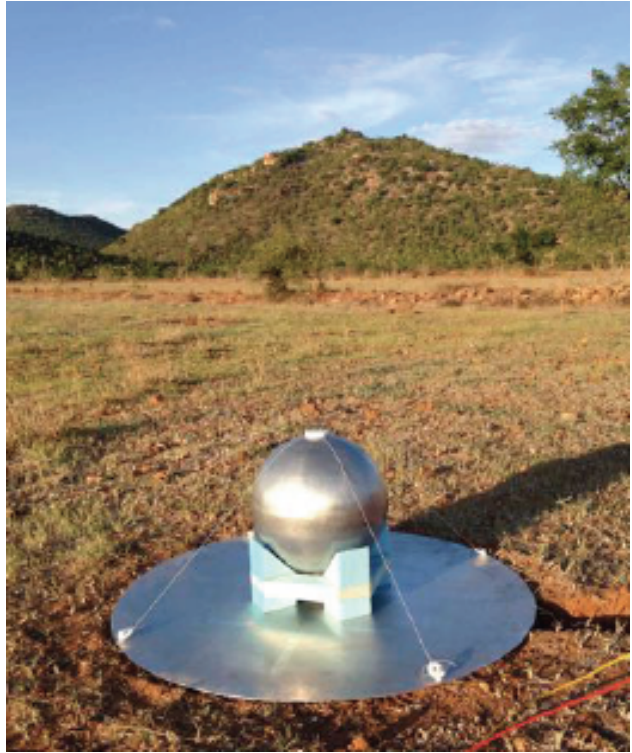

Figure 6-3 SARAS-2 Sphere-disc antenna

### 6.1.3. Sci-HI and High-z experiment

The Sonda Cosmologica de las Islas para la Deteccion de Hidrogeno Neutro (SCI-HI) experiment makes use of matched modified four-point antenna called a Hibiscus Antenna. In 2014 the team published few-Kelvin-precision upper limits to possible first-stars spectral structure (Peterson et al. 2014, Voytek et al. 2014). Since that time the spectrometer has been rebuilt with a higher precision Analog to Digital Convertor, and the experiment now uses an updated antenna design called Mango-Peel.

The experiment is operated on Isla Guadalupe in the Pacific Ocean 165 km west of Ensenada, Mexico. The west coast of the island, shielded from the mainland by the North-South ridgeline at elevation about 1000 m, offers very low FM-band RFI. The island is accessible by boat or by small aircraft, offering easy accessibility, but access to the island requires a government permit, so development is limited. A solar-charged battery system has been procured for use on the island.

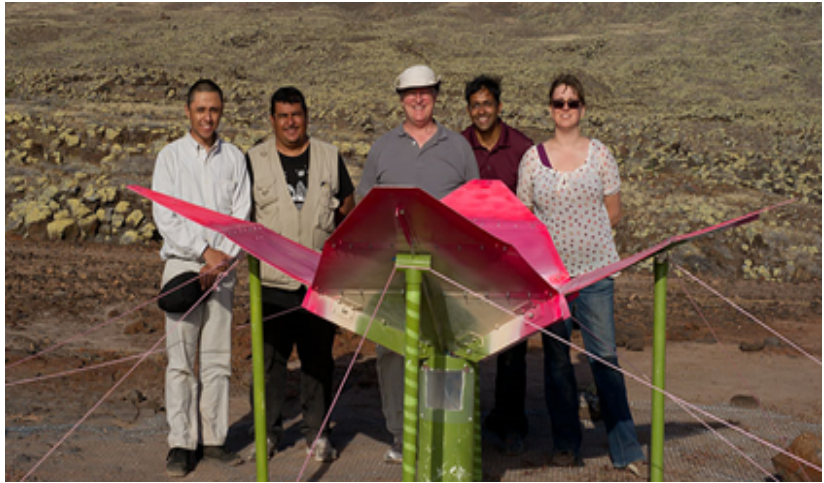

Figure 6-4. The SCI-HI Hibiscus antenna on Isla Guadalupe.

Using a high impedance antenna, connected to an even higher impedance amplifier, the High-Z all-sky spectrometer is designed to measure the sky-averaged brightness from 25 to 200 MHz, constraining the cosmic dawn era of cosmology. This three-octave range offers a substantial improvement over the one-octave range of previous efforts, allowing better control of ionospheric effects in the data while also permitting any potentially-detected cosmological signals to stand out statistically from reference spectral regions. The use of very high impedance readout eliminates the need for field measurement of the antenna coupling efficiency and is anticipated to reduce the error associated with this coupling factor.

High-Z uses an active monopole antenna: an unloaded electrically-short vertical antenna. By reading the antenna voltage without significant resistive current flow, a calibrated flat multi-octave response to the electric field is achieved. Unlike previous all-sky experiments there is no need to measure antenna power coupling efficiency. Indeed, for the High-Z system this quantity is nearly zero. Instead of matching the antenna, High-Z makes use of an amplifier with parallel resistive impedance orders of magnitude higher than the antenna. In this limit the actual value of the amplifier impedance is irrelevant: the amplifier simply measures the undisturbed antenna voltage. In addition to offering reduced matching uncertainty, the unloaded antenna allows for a multi-octave passband since the antenna factor becomes frequency independent at long wavelengths.

No other Global 21-cm experiment uses this ultra-wide-band scheme. The SARAS-II experiment uses a short vertical antenna, but reads the antenna using a 50-ohm amplifier. This limits the useful bandwidth to about one octave. All other current experiments use compact matched antenna designs that are again limited to about one octave frequency ranges. The High-Z antenna offers a flat antenna factor from zero frequency to an upper limit defined by the quarter-wave resonance in the vertical antenna. The three octave range is set not by the antenna but by the high and low pass filters within the spectrometer.

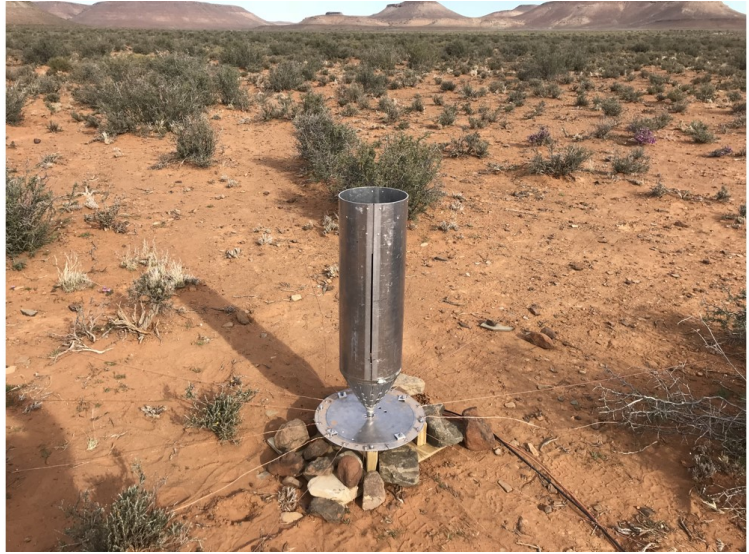

Figure 6-5 The 84 cm antenna deployed at Karoo Site 4 in July 2018. Ground radial wires attach the center plate. The calibration system and LNA are mounted in the RF box below this plate.

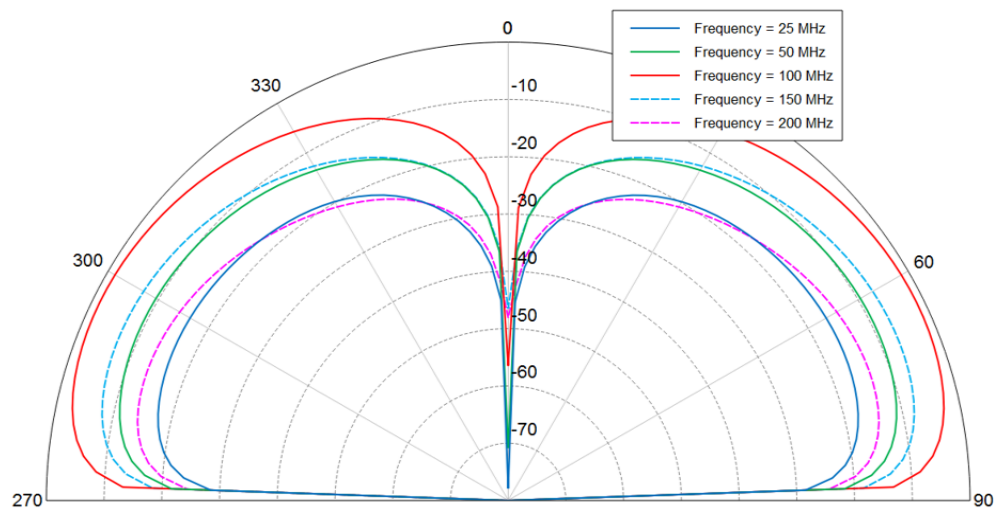

Figure 6-6 **Numerically calculated antenna patterns.** Patterns for the Medium antenna are shown in a vertical plane, and are independent of azimuth as expected from the antenna symmetry. Pattern shapes are very similar across the entire three-octave range. The broad 1/4 wave resonance is evident near 100 MHz, but this has very little effect on the pattern. The pattern shape is free of multi-lobe structure across three octaves.

In order to study the effect of antenna length on the measured sky brightness the team built an antenna with removable extensions, allowing three lengths: 25 cm, 54 and 84 cm. The 25 cm antenna is the primary antenna to be used for cosmological observations since its quarter-wave resonance lies above the 200 MHz top edge of the High-Z passband. The other antennas each have an in-band resonance. These antennas are used to examine the width and shape of the quarter-wave resonance peak, including any low frequency wing of the antenna factor resonance curve that may affect the system response near the upper end of the pass band when using the short antenna. The antenna also uses a horizontal ground plane, consisting a 38 cm aluminum disc, connected to 16 ea. 1 mm diameter stranded radial wires of length 40m. The radial wires are uninsulated and rest on the soil. Figure 6-5 shows a photo of the 84cm antenna.

The High-Z system is fit into three large suitcases and can be set up at a remote site in a few hours. Solar-charged batteries power the system so it can be used at sites that lack AC power. The lack of AC power eliminates possible sources of local RFI. High-Z has been tested in the field at Green Bank West Virginia, Algonquin Radio Observatory and twice at the Karoo Astronomy Reserve in South Africa. Figure 6-5 shows the antenna deployed at Karoo. Figure 6-6 shows calculated antenna patterns.

The high impedance antenna used in High-Z may also be useful in a spectrometer in lunar orbit or on the lunar surface. The small size and ultra wide bandwidth of these designs is well suited the frequency range accessible from the moon.

### 6.1.4 PRI$^Z$M/ALBATROS

Probing Radio Intensity at high-Z from Marion (PRI$^Z$M) is an experiment that has been specifically designed to study cosmic dawn in the universe using 50-200 MHz observations of globally averaged 21-cm emission. The experiment consists of two compact, modified four-square antennas that operate at central observing frequencies of 70 and 100 MHz. One of the primary challenges for low-frequency radio observations is terrestrial radio-frequency interference (RFI), which can overwhelm the cosmological signal even when the nearest RFI sources are hundreds of kilometers away. PRI^ZM complements other cosmic dawn experiments by observing from Marion Island, which lies roughly 2000 km from the nearest continental land masses and is one of the most isolated locations on Earth. PRI^ZM was first installed on Marion in 2017, and a complete description of the instrument is available in Philip et al., (2019). Our preliminary observations demonstrate that Marion provides an exceptionally clean RFI environment, with essentially no visible RFI within the FM band. Remarkably, Marion even outperforms the Square Kilometre Array site in the Karoo desert, which is one of the premier radio astronomy observing locations worldwide.

We are in the early phases of extending our observing program from Marion Island to even lower frequencies. At frequencies around tens of MHz, future observations may allow us to one day probe the cosmic "dark ages," an epoch that is unexplored to date and contains a potential wealth of information, as described elsewhere in this report. Measurements at these frequencies are extremely challenging because of RFI contamination and ionospheric effects; the combination of Marion's clean RFI environment and the 2019 solar minimum presents a unique opportunity to perform exploratory low-frequency measurements. The Array of Long Baseline Antennas for Taking Radio Observations from the Sub-antarctic (ALBATROS) is a new interferometric array that is under development for installation on Marion. The final array will comprise roughly 10 autonomous antenna stations operating at 1.2-81 MHz (with baseband recorded for the lowest ~10 MHz) and with baselines up to 20 km. A two-element pathfinder was installed in April 2018, and the preliminary cross-correlation data show repeatable interference fringes from the sky down to about 10 MHz without any processing or cuts. We will continue to expand the ALBATROS installation on Marion in future voyages.

## 6.2 Ground Imaging Experiments

### 6.2.1. LOFAR

The Low-Frequency Array (LOFAR; van Haarlem et al. 2013) has the detection

of the 21-cm signal of neutral hydrogen during the Epoch of Reionization (EoR; see, e.g. Patil et al. 2017) as one of its key-science drivers. LOFAR is also a pathfinder to the Square Kilometre Array (in particular SKA-low). The few-km inner core of LOFAR High Band Antenna (HBA; 110-240MHz) system consists of 48 stations each having 24 tiles of each four by four cross-dipoles. The large collecting area in the core and the high filling factor of LOFAR-HBA make it particularly suited for the detection of diffuse low-surface brightness emission of neutral hydrogen during the EoR. Besides LOFAR observing in standard station beam-forming mode, it can also cross-correlate all 576 HBA tiles or LBA dipoles (see below) of the inner 12 stations, yielding an extremely large field of view (e.g. Gehlot et al. 2018a&b) and hence probe a larger volume of the Universe, reducing sample variance that dominate the largest scales of the 21-cm signal. The Low-Band Antenna system (i.e. (10)-30MHz) could, in principle, detect the 21-cm signal from the Cosmic Dawn (CD) as well (e.g. Gehlot et al. 2018a&b), but only if the 21-cm signal exceeds nominal predictions of 10-100 mk2 at k=0.1 cMpc-1 by a significant fraction (e.g. Fialkov et al. 2018, 2019). The reason is the extremely bright Galactic foregrounds that cause a high noise level. The second reason is that ionospheric effects are far more significant at lower frequencies, causing signal distortions.

Currently, LOFAR has yielded the deepest upper limits on the 21-cm signal from both the EoR and the CD (Patil et al. 2017, Gehlot et al. 2019b) with a publication on an ever deeper limit in preparation. Although PAPER claimed a deeper upper limit, this result has recently been retracted due to signal suppression in the processing, something that LOFAR can control thanks to its longer baselines (Mouri Sardarabadi & Koopmans 2018).

The lessons to be learned from LOFAR, for DSL, are foremost that signal suppression in the imaging and calibration process needs to be taken seriously and needs to be mitigated via either sophisticated signal-processing algorithms or via special observational calibration schemes, in order to obtain reliable results. Both the sky and the system model need to be obtained from the same data set, and signals can leak from one to the other (i.e. gain errors can remove sky brightness, or add it, and visa versa). Secondly, at very low frequencies (<50MHz) the ionosphere is a serious issue that will ultimately limit the dynamic range in any observation. DSL will not suffer from this. Hence probing the Dark Ages (z>40, or ν <35MHz) will most likely only ever be successful from space. Although RFI has a limited impact on LOFAR (and other ground-based instruments), it is an increasing problem, that might require most future experiments to be conducted from space.

One recommendation from the LOFAR experience would be to consider having both short and long baselines for DSL, the latter being used to help build sky models on which the entire data-set can be calibrated. The long baselines (>>few hundred meters) are not sensitive to the 21-cm signal, but do see compact sources on which calibration is much easier. The very short baselines are needed for the 21-cm signal detection (signal fluctuations), but DSL will likely not have the collecting area for this. The global 21-cm signal, on the other hand, can be obtained from signal autocorrelations, but might still require a sky model in order to mitigate spectral distortions in the signal due to a frequency dependent beam of the receiver coupling to the sky. DSL itself, as well as ground-based instruments such as LOFAR can provide such sky models, making them complementary.

### 6.2.2. MWA

The emissive and reflective properties of the Moon are important for the DSL mission as an accurate model of the Moon will need to be included in the sky model used for calibration. These properties will affect both the imaging and global-signal components of the mission and are not well-known at low radio frequencies, even for the near side of the lunar surface. The Moon has been imaged at low frequencies by both LOFAR and the Murchison Widefield Array (MWA; Tingay et. al., 2013). The MWA observations (McKinley et al. 2013, 2018) cover the frequency range 72 - 230 MHz. This range overlaps with the frequency coverage of the DSL global experiment (30 - 100 MHz), but is still significantly higher than the imaging component (1-30 MHz). MWA observations, however, may still be useful for determining some of the physical characteristics of the Moon that give rise to these properties. For example, the high signal-to-noise ratio afforded by observing lunar reflections in the FM band may be able to be used to measure the ripples in the signal due to interference between waves reflected at the topsoil surface, and a layer of solid bedrock at some distance below the lunar regolith, as early LOFAR results in the high-band have hinted at. By learning as much as possible about the Moon's radio properties with ground-based instruments such as the MWA and LOFAR, we can give the DSL mission a reasonable model to be used for initial calibration.

## 6.3 Space Experiments

### 6.3.1 SUNRISE

Our Sun produces intense storms of high energy particles that flood interplanetary space. These storms are of high scientific interest because we do not understand how nature is able to accelerate particles in the solar atmosphere to such high energies so quickly and efficiently. They are also major contributors to space weather, able to disable and even destroy instruments and spacecraft, and posing a potentially lethal threat to unprotected astronauts in deep space or on the lunar or Martian surface. The Sun Radio Interferometer Space Experiment (SunRISE) fills a major knowledge gap in the connection between the physics of particle acceleration and the production and transport of energetic particles from the Sun into interplanetary space ([Lazio et al., 2017](#)). Particle acceleration is one of the most pressing questions in Heliophysics. Radio frequency observations provide a well-established remote sensing technique for probing sites of particle acceleration, but Earth's ionosphere distorts and absorbs the low frequency radio waves below 15 MHz produced by energetic particles as they move more than a couple solar radii away from the Sun. Orbiting just past geosynchronous orbit and far above the ionosphere, SunRISE will be the first mission able to image and track the telltale radio emissions. Launch is planned for 2022.

At these low frequencies a classical focusing radio dish would be 10 km across – very impractical to launch, assemble, or even point in space. SunRISE eliminates the need for a physical dish by forming a virtual one. SunRISE consists of six CubeSat-sized spacecraft flying in a 10 km diameter constellation. As a radio interferometer, SunRISE records radio waves so precisely that the effects of a dish are reproduced by delaying and combining the waves seen at each spacecraft to focus on any desired location in the sky. Interestingly, SunRISE data can be recombined on the ground to repoint the array after the fact in any direction of interest.

Many objects of interest in the solar system, including coronal mass ejections and Earth's high energy radiation belts, produce intense radio emission at low frequencies, below 15 MHz, that are blocked by Earth's ionosphere, especially during solar flares. The location and frequency of the emission reveals essential information about where energetic particles are created, stored, and how they move.

SunRISE will make the first images of the location and motion of intense solar radio bursts associated with particle acceleration by coronal mass ejections (CMEs) and solar flares. The occurrence of these bursts is well established, with observations of their intensity as a function of frequency and time commonly recorded by antennas on single spacecraft since the beginning of the space age. We know for example that every CME that produces a major solar energetic particle (SEP) event at Earth first emits an intense Type II radio burst starting about 10-20 minutes before the SEPs are released into space. The first objective of SunRISE is to image where these Type II bursts are relative to the erupting CME to determine what aspects of the CME lead to intense acceleration. We also know that flares release energetic particles into space, and that these particles generate intense Type III radio bursts that rapidly drop in frequency as the particles move into space. The second objective of SunRISE is to map the path taken by particle radiation from the Sun into space by making the first images of Type III bursts.

### 6.3.2. DARE and DAPPER

Burns et al. (2017) have proposed an Explorer-class mission, the Dark Ages Radio Explorer (DARE), to make space-based observations of the 21-cm monopole from a low lunar orbit. The DARE instrument design incorporates (1) an optimized antenna with on-orbit beam calibration, (2) the replacement of Dicke switches for bandpass calibration with a pilot frequency tone system capable of high dynamic range monitoring of gain variations and measurements of the system reflection coefficients, and (3) polarimetric observations to provide a model-independent measure of the beam-averaged foregrounds. The observations, performed from the radio-quiet zone above the Moon's farside, will be enabled through a unique "frozen" 50×125 km lunar equatorial orbit. The nominal observation noise integration is to the 1.7 mK level at 60 MHz. The instrument provides data with the frequency range (40–120 MHz), spectral resolution (50 kHz), beam FWHM at 60 MHz), and polarization required to measure the spectral features expected from the wide range of theoretical models considered. DARE's observing strategy utilizes four quiet-sky pointing directions away from the galactic center.

More recently, Burns et al. (2019) proposed a SmallSat low frequency experiment called the Dark Ages Polarimeter PathfindER (DAPPER) to fly in conjunction with NASA's accelerated lunar exploration program. DAPPER is proposed to observe at frequencies 17-38 MHz (z~83-36), and will measure the amplitude of the 21-cm spectrum to the level ) the standard cosmological model from that of additional cooling derived from current EDGES results. DAPPER's science instrument consists of dual orthogonal dipole antennas and a tone-injection spectrometer/polarimeter based on high heritage components from the Parker Solar Probe/FIELDS, THEMIS, and the Van Allen Probes. DAPPER will be deployed from the vicinity of NASA's Lunar Gateway or in cis-lunar 125 km lunar orbit using a deep-space spacecraft bus that has both high impulse and high delta-V. This orbit will facilitate the collection of 4615 hours of radio-quiet data over a 26.4 month lifetime.
The early Universe's Dark Ages, probed by the highly redshifted 21-cm neutral hydrogen CDM cosmological model.

DAPPER will search for divergences from the standard model that will indicate new physics such as heating or cooling produced by dark matter. The Cosmic Dawn trough is affected by the complex astrophysical history of the first luminous objects. Another trough is expected during the Dark Ages, prior to the formation of the first stars and thus determined entirely by cosmological phenomena (including dark matter). DAPPER will observe this pristine epoch (17-38 MHz; z~83-36), and will measure the amplitude of the 21-cm spectrum to the level required to distinguish ) the standard cosmological model from that of additional cooling derived from current EDGES results. In addition to dark matter properties such as annihilation, decay, temperature, and interactions, the low-frequency background radiation level can significantly modify this trough. Hence, this observation constitutes a powerful, clean probe of exotic physics in the Dark Ages. A second objective for DAPPER will be to verify the recent EDGES results for Cosmic Dawn, in the uncontaminated environment above the lunar farside, with sparse frequency sampling from 55-107 MHz (z∼25-12). The combination of both troughs will robustly augment the constraining power of DAPPER on early-Universe astrophysics and cosmology.

The main challenge of this measurement is the removal of bright foregrounds. DAPPER is designed to overcome this by utilizing two techniques: (1) a polarimeter to measure both intrinsically polarized emission and polarization induced by the anisotropic foregrounds and large antenna beam to aid in the separation of the foregrounds from the isotropic, unpolarized global signal, and (2) a pattern recognition analysis pipeline based on well-characterized training sets of foregrounds from sky observations, instrument systematics from simulations and laboratory measurements, and signals from theoretical predictions. DAPPER team members recently demonstrated the effectiveness of dynamic polarimetry to measure foregrounds using observations from the prototype Cosmic Twilight Polarimeter. Rigorous end-to-end simulations of the DAPPER instrument including thermal noise, systematics from the spectrometer/polarimeter and the beam-averaged foreground, along with 21-cm models which include added cooling meet our sensitivity requirements to separate the standard cosmological models from ones that point toward new physics.

DAPPER's science instrument consists of dual orthogonal dipole antennas and a tone-injection spectrometer/polarimeter based on high heritage components from the Parker Solar Probe/FIELDS, THEMIS, and the Van Allen Probes. DAPPER will be deployed from the vicinity of NASA's Lunar Gateway or in cis-lunar 125 km lunar orbit using Bradford Space Industries Xplorer deep space bus which has both high impulse and high delta-V. This orbit will facilitate the collection of 4615 hours of radio-quiet data over a 26.4 month lifetime for the baseline mission.

DAPPER is a collaboration between the universities of Colorado-Boulder and

California-Berkeley, the National Radio Astronomy Observatory, Bradford Space Inc., and the NASA Ames Research Center.

### 6.3.3. PRATUSH

The Indian Space Research Organization (ISRO) has provided pre-project funding for detailed design of a lunar orbiter mission PRATUSH, Probing ReionizATion of the Universe using Signal from Hydrogen, proposed by PIs Mayuri S Rao, Saurabh Singh and Jishnu Nambissan T from the Raman Research Institute, India.

# 7. Potential Issues

## 7.1 EMI suppression and RFI removal

Although space-borne radio astronomy instruments may face much lower levels of man-made radio interference (RFI) than Earth-based instruments, interference generated by the spacecraft itself needs to be carefully considered when designing small scientific radio satellites. Traditionally, low-frequency space and plasma science operates roughly between 30 kHz and 15 MHz, and typically relatively narrow spectral receiving bands are used to help minimize the impact of RFI. Space-based radio astronomy aims at the regime 1-30 MHz and also at higher frequencies to allow cross-calibration with Earth-based radio telescopes. The science cases push for broad-band observations, and technologically this is becoming feasible. Operating in broad bands however makes receiver systems more vulnerable to RFI, especially in terms of linearity. For this reason it usually is better to split the band into a few analogue subbands, although the digital electronics is capable enough to sample the entire band in one go. Both planetary AKR signals and self-generated RFI can be very strong, especially at frequencies around and below 1 MHz.

Self-generated RFI from spacecraft systems and from scientific receiver equipment is best tackled in the design phase of both, as reducing these emissions in existing designs is very difficult and costly. Apart from choosing an adequate grounding and shielding philosophy, special care must be taken to limit emissions from spacecraft clocks and power converter units. Choosing narrow scientific observational frequency channels (down to kHz level, if the system budget allows this) is advantageous as this would allow excising narrow-band RFI.

In terms of algorithmic RFI mitigation approaches, there are many ways to detect and suppress RFI. Intermittent RFI can be detected by using a power detector or higher order statistics (kurtosis, cyclo-stationarity), although the later may require significant computing resources, and may not be available on-board. For multiple antennas on a spacecraft (two or three orthogonal dipoles polarization, or a number on monopole antennas) one can also use spatial filtering to suppress RFI. It may be advantageous to add an additional (small) reference antenna. As it is possible to find a direction of arrival (DoA) with just one set of co-located antennas (using goniopolarimetry), it is also possible to suppress signals from a certain DoA. This principle, to some extent, is also applicable to RFI generated by the spacecraft itself.

An array of radio satellites to be used in interferometric mode in principle can apply the same suite of temporal, spectral, and spatial RFI mitigation options that ground based telescopes have. Main limiting factor is computational load. It is worth to consider off-loading part of the data processing to the calibration and scientific processing on Earth. For example, if the radio array produces narrow-band correlation data, then it is more efficient to apply spatial filters in the post processing step.

## 7.2 Calibration

For the imaging array, the observing frequencies of between 1-30 MHz have never been imaged before at sub-degree resolution. This represents a unique challenge for calibration, since there is no known source in the sky that we can use to calibrate the instrument. In order to construct a sky model for calibration we must rely on the few low-angular-resolution maps at these low frequencies made from the ground (e.g. Caswell, 1976 at 10 MHz; Roger et. al., 1999 at 22 MHz), theoretical estimates of the various emission and absorption mechanisms and extrapolations from higher frequencies (see e.g. Huang et. al., 2019 for an example of a self-consistent sky model).

## 7.3 Imaging

Each of the daughter satellites will be equipped with three, orthogonal dipole antennas. By the very nature of these antennas, the imaging array will be sensitive to the full, 4π steradians of the sky. In any other imaging array there is always a ground-screen (either part of the instrument or the Earth itself) that restricts the field-of-view to, at most, half of the sky. Various methods can then be used to project the spherical sky onto a flat surface to be represented as an image. For telescopes with a small field-of-view the sky can be approximated as a 2-D surface and images can be made simply by Fourier transforming the calibrated visibilities. The ultra-long wavelength lunar-orbiting array will not be able to make any such approximations and will need to take into account the full sky. This means that sources of emission will be in view both 'in front of' and 'behind' the array. Conventional radio astronomy imaging algorithms are unable to distinguish between the two sides, resulting in the 'mirror symmetry' problem whereby sources from behind the array appear as if they are in front (Huang et. al., 2018). Therefore, unconventional methods, such as brute-force imaging (Zheng et. al., 2017, Huang et. al. 2018a) will need to be trialled and new, bespoke imaging algorithms may need to be developed. These are likely to be very computationally expensive as they will need to make use of the full 3-D distribution of *u,v,w* points in order to overcome the mirror symmetry problem.

The DSL array is in constant motion with an orbital period of 2.3 hours. This means that the u,v,w coordinates are changing much faster than a ground-based array. Hence, for the imaging array, the integration times must be very short to avoid time decorrelation (38 ms), which is baseline degradation of the data due to averaging over time. Also, unlike a ground-based array where the coordinates of the antennas can be measured to extremely high precision using GPS and the baselines lengths always remain constant, the baseline lengths of the space-based array will be changing and will be known to 1 m precision through the use of the range determination system.

**7.4 Foreground Subtraction**

At an orbital altitude of 300 km, the Moon will take up about π steradians of the sky. The part of the sky obscured by the Moon will be constantly changing over the course of the orbit. Sky models for calibration will need to take into account this obscuration, but it will not be a simple mask function. The Moon has intrinsic thermal emission and also reflects radio waves ([McKinley et. al. 2018](), [Vedantham et. al. 2015]()). These reflections will be dominated by Galactic synchrotron emission, which is expected to be extremely bright at these ultra-low frequencies. With baselines as short as 100 m and frequencies as low as 1 MHz, the imaging array will be sensitive to the angular scales of the Moon's emission and reflection. Hence the thermal and reflective characteristics of the Moon (which require further study at low radio frequencies) must be taken into account in simulations and calibration. For the global-signal experiment, the Moon's emission and reflections may leak into the antenna through sidelobes and hence it is important to simulate these effects using antenna beam models and our best approximation of the properties of the Galactic sky and the Moon.

## 8. Summary

During this forum we have discussed various aspects of the low frequency radio astronomy, including ground based imaging array, global spectrum measurement, and particularly space-borne experiments with a small satellite constellation. The ultra long wavelength radio spectrum has great potentials for scientific breakthrough, especially for the study of the cosmic dawn and dark age. Imaging the neutral hydrogen during the dark age can provide information about the inflationary origin of the Universe, but this requires extremely large arrays on the far side of the Moon. A useful step first is to map the foreground which is necessary for the design of future dark age experiments, and high precision global spectrum may enable us to make a first peek at the cosmic dawn. The ultra long wavelength observations may also bring interesting discoveries in solar burst, planets and exo-planets, the interstellar medium and intergalactic medium, evolution of extragalactic radio sources, cosmic ray and ultrahigh energy neutrinos.

However, for the limited resource of the first attempts, it is very important to focus on the most important science cases and obtain tangible results.

DSL is a lunar orbit mission for low frequency radio astronomy. As discussed in earlier sections, space-borne observation can overcome the problem of strong distortion and absorption by ionosphere at frequencies below 30 MHz, and the lunar far side provides an ideal place where the radio emission from the Earth can be shielded. Compared with observation from the lunar surface, the lunar orbit observation is not directly affected by the Sun-induced ionosphere on the day-side lunar surface, and furthermore, there are also a few advantages from an engineering point of view: (i) A lunar orbit mission does not need to land on the moon, so there is no need for the additional landing system. (ii) The low lunar orbit period is less than a few hours, the power can be supplied with conventional solar-panel and battery, whereas a lunar lander must deal with long lunar nights. (iii) The data can be transmitted back to Earth during the time when the satellite is at the near side part of the orbit, there is no need to have special relay satellites. All of these points mean that a lunar orbit mission is less complicated and less expensive than a lunar surface mission of similar capacity.

The DSL has a very exciting science case and also a plausible technological concept. However, there remains many technological challenges and uncertainties. International collaboration would enrich and advance these researches. The coming decade offers a unique opportunity for RF-quiet measurements from the lunar far side, before further development of lunar assets compromises the RF-quiet character in the long-term.

## Acknowledgement

The present paper is based on the report of the International Space Science Institute in Beijing (ISSI-BJ) forum on "Discovering the Sky by the Longest Wavelengths with Small Satellite Constellations", which was held at ISSI-BJ on 23rd-25th January 2019. We thank the staff members of the ISSI-BJ for providing logistic supports. The DSL project is supported by the Chinese Academy of Sciences Strategic Priority Research Program XDA15020200.